\documentclass[english,aps,jcp,showpacs,twocolumn]{revtex4-1}
\usepackage[T1]{fontenc}
\usepackage[latin9]{inputenc}
\usepackage{amsmath}
\usepackage{graphicx}
\usepackage{amssymb}
\usepackage{color}
\usepackage{hyperref}
\renewcommand{\vec}[1]{\mathbf{#1} }
\usepackage{babel}

\begin{document}

\title{Monte Carlo computer simulations and electron microscopy of colloidal cluster formation via emulsion droplet evaporation}

\author{Ingmar Schwarz}

\author{Andrea Fortini}

\author{Matthias Schmidt}
\email{Matthias.Schmidt@uni-bayreuth.de}
\affiliation{ Theoretische Physik II, Physikalisches Institut, Universit\"at Bayreuth, Universit\"atsstra{\ss}e 30, D-95440 Bayreuth,
Germany}

\author{Claudia Simone Wagner}
\author{Alexander Wittemann}

\affiliation{ Physikalische Chemie I, Universit\"at Bayreuth, Universit\"atsstra{\ss}e 30, D-95440 Bayreuth,
Germany}

\begin{abstract}
We consider a theoretical model for a binary mixture of colloidal particles and spherical emulsion droplets. The  hard sphere colloids interact via additional short-ranged attraction and long-ranged repulsion. The droplet-colloid interaction is an attractive well at the droplet surface, which induces the Pickering effect. The droplet-droplet interaction is a hard-core interaction. The droplets shrink in time, which models the evaporation of the dispersed (oil) phase, and we use Monte Carlo simulations for the dynamics. In the experiments, polystyrene particles were assembled using toluene droplets as templates. The arrangement of the particles on the surface of the droplets was analyzed with cryogenic field emission scanning electron microscopy. Before evaporation of the oil, the particle distribution on the droplet surface was found to be disordered in experiments, and the simulations reproduce this effect. After complete evaporation, ordered colloidal clusters are formed that are stable against thermal fluctuations. Both in the simulations and with field emission scanning electron microscopy, we find stable packings that range from doublets, triplets, and tetrahedra to complex polyhedra of colloids. The simulated cluster structures and size distribution agree well with the experimental results. We also simulate hierarchical assembly in a mixture of tetrahedral clusters and droplets, and find supercluster structures with morphologies that are more complex than those of clusters of single particles.
 \end{abstract}

\maketitle

\section{Introduction}
Systems of colloidal particles that interact with pair potentials that feature short-ranged attraction have been extensively studied in the literature. The typical bulk phase diagram shows a characteristic  gas-liquid phase transition that can become metastable with respect to freezing for  sufficiently short range of attraction~\cite{Lekkerkerker1992,Ilett1994,Schmidt2000,Dijkstra1999a}. 
%A variety of fundamental many-body phenomena such as wetting of substrates\cite{Wijting2003a,Wijting2003,Aarts2004a,Wessels2005,Evans2003,Dijkstra2002,Brader2002,Binder2003}, capillary waves at fluid interfaces~\cite{Aarts2004,Derks2006,Vink2005c}, glass transition and gelation~\cite{Segre2001,Sciortino2002,Pham2002,Puertas:2007p890,Puertas:2005p858,Manley2005,Zaccarelli2007,Sciortino2005a,Lu:2008,Lodge1999,Cates2004a,Candia2005,Bergenholtz2003,Fortini:2008a} have been investigated.
While the addition of a short-ranged soft repulsion is known to change  the phase diagram~\cite{Fortini2005,Pini:2006p279} only quantitatively,  a repulsion with a range larger than that of the attraction can have a dramatic effect on the topology of the phase diagram~\cite{Candia2006}.  
Moreover, both in experiments and in dynamical simulations the equilibrium phase diagram for systems with competing interactions is often overshadowed by non-equilibrium phenomena such as vitrification, gelation and cluster formation~\cite{Sciortino2005,Mossa2004,Campbell2005,Coniglio2006,Archer:2007p4145}.
Typically, pair potentials with a short-ranged attraction show spontaneous clustering of particles at sufficiently low temperatures.
The geometric structure of clusters  of particles has been studied theoretically, e.g. for the Lennard-Jones potential~\cite{HOARE:1971p3888,Taffs:2010p3894}, hard spheres~\cite{SLOANE:1995p3886}, and  hard spheres that additionally interact with a short-ranged attraction~\cite{Arkus:2009p3887}. The structures obtained using the Morse potential were analyzed by~\citet{Doye:2011p3893} and by~\citet{Taffs:2010p3892} and compared to those  from the Asakura-Oosawa potential~\cite{Asakura1954,Vrij1976}.
Stable clusters of colloids are  interesting because they can be viewed as colloidal molecules~\cite{Blaaderen2003,duguet2011} that can potentially be used as building blocks for the  fabrication of novel materials.  
The  size and geometric structure of colloidal clusters, however, are not easily controllable in experiments. Several different methods that offer control of the clustering process have been proposed. 
 \citet{Jiang:2007p902} prepared clusters using Janus colloidal particles, i.e. spherical particles that possess oppositely charged hemispheres. Cluster of particles with larger numbers of patches have also been studied~\cite{Zhang:2004p3779,Wilber:2007p4113,Wilber:2009p2272}. \citet{Erb:2009p703} succeeded in preparing clusters of magnetic particles.

%emulsions
One particularly promising approach, based on the evaporation of the dispersed phase in an emulsion, 
 was developed~\cite{Velev:1997p909} by~\citet{Velev:1996p4063,Velev:1996p906}. Here colloidal particles adsorb at the interface between dispersed and continuous phase in order to minimize the interfacial free energy (Pickering effect)~\cite{binks}.
During evaporation the particles are pushed together by capillary forces and subsequently held together by van der Waals interactions. 
 \citet{Manoharan:2003p937} prepared micron-sized clusters using this technique with  polystyrene microspheres that were 844 nm in diameter. The authors found clusters of particles with packings that minimize the second moment of the mass distribution. The emulsion method is versatile and was subsequently used to obtain clusters of particles 220 nm in diameter~\cite{Cho:2005p916}, of  patchy particles~\cite{Cho:2007p814,Kim:2008p851}, and of bidisperse colloids~\cite{Cho:2005p908}. Shear was used by \citet{Zerrouki:2006p4273} to produce monodispersed droplets. A similar technique is based on aerosol droplets~\cite{Cho:2007p919}  instead of oil droplets.

%miniemulsion
Similar in spirit to the emulsion evaporation technique, an alternative miniemulsion technique~\cite{Wagner:2008p844,Wagner:2010p3885} was recently developed.  Here, a miniemulsion is prepared from a standard emulsion by ultrasonication. The sound waves produce an emulsion of small droplets in a process of fission and fusion [65]. The average size of the droplets can be tuned in the range of 360 nm to 1800 nm~\cite{Landfester:2000p1556}. 
Small colloidal particles can then be used in solution with the small monodispersed droplets of the miniemulsion to obtain clusters~\cite{Wagner:2010p3885} that have diameters much smaller than 1 $\mu m$. Although these clusters can consist of many constituent colloidal spheres, they remain small enough to reside in the colloidal domain~\cite{Hoffmann:2009p2493}, i.e. Brownian motion thermalizes such systems. Therefore hierarchical self-assembly comes within reach.   

In contrast to the large body of experimental work and the closely related theoretical efforts to understand the resulting cluster structures and their symmetries, little theoretical work has been done to describe the {\it process} of cluster formation. 
\citet{Roman:2000p4109} proposed a model for a dispersion of hard spheres and emulsion droplets, but  these authors did not  investigate  cluster formation. 
\citet{Lauga2004}  modeled and simulated the evaporation-driven
   assembly of colloidal particles. They considered individual
   droplets with varying numbers of adsorbed particles and calculated
   the (non-spherical, in general) shape of the oil-water interface by
   the requirement of minimal surface free energy. They considered
   different values of the contact angle and reported good agreement
   with experimental findings.
Very recently, \citet{Mani:2010p4140} studied the stability of larger colloidosome-like shells of particles, albeit without modeling the assembly process.

In this paper we present a basic model to describe the process of cluster assembly through emulsion droplet evaporation. We use Monte Carlo computer simulations to study the cluster formation of colloids with competing short-ranged attraction and long-ranged repulsion interactions. 
Differently from \citet{Lauga2004} we also simulate the dynamical capture of the colloidal particles onto droplet surfaces and studied not only the cluster structures, 
but also analyzes the histograms of the cluster size distribution.
We complement these calculations with experiments of polystyrene particles in an oil-in-water emulsion. 
We use cryogenic field emission scanning electron microscopy (cryo-FESEM) and field emission scanning electron microscopy (FESEM) to investigate the intermediate and final stages of cluster formation, respectively. The cryo-FESEM micrographs show the distribution of small particles on a droplet surface.
We find the simulation results to be in good agreement with the experimental results for both intermediate and final cluster structures. We also find  good quantitative agreement between experimental  and simulation results for the cluster size distribution.
Having demonstrated the validity of our model, we study the possibility of hierarchical self-assembly, by carrying out simulations  of a mixture of thermal tetrahedral clusters and emulsion droplets. We obtain clusters of clusters (superclusters) with  structures that differ from the clusters made of single particles.

The paper is organized as follows. In Sec.~\ref{s:int} we give the details of the pair interactions. In Sec.~\ref{s:sim} and \ref{s:exp} we present simulation and experimental details, respectively.  In Sec.~\ref{s:fluid}, we describe the results for the dynamics of cluster formation. In Sec.~\ref{s:cls}, we show the results for the cluster structures and for the histograms of the size distributions. In Sec.~\ref{s:super}, we present the results for the superclusters.
Final remarks and conclusions are given in Sec.~\ref{s:conc}.

\section{Model and Methods}

\begin{figure}
\includegraphics[width=8cm]{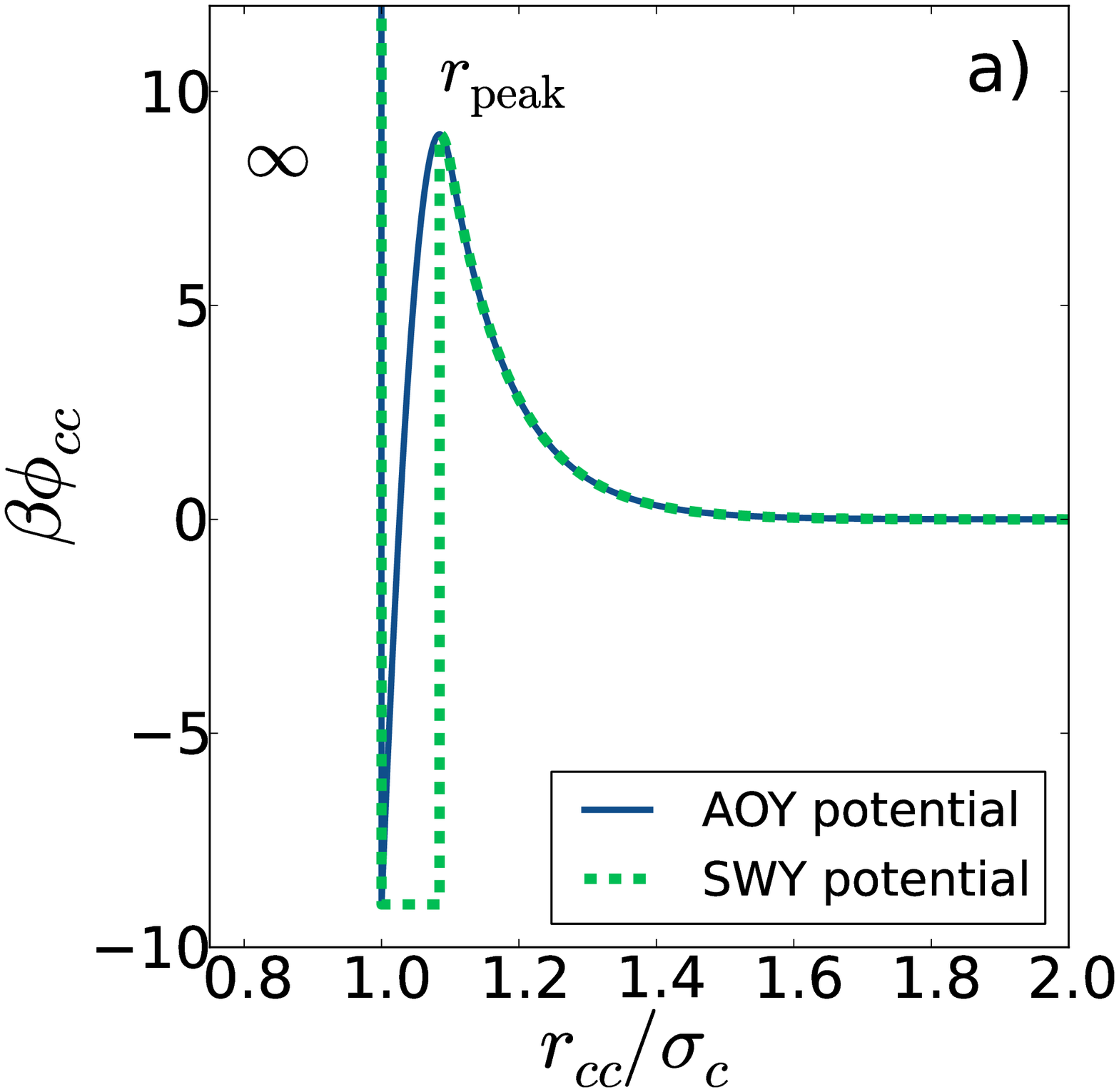}
\includegraphics[width=8cm]{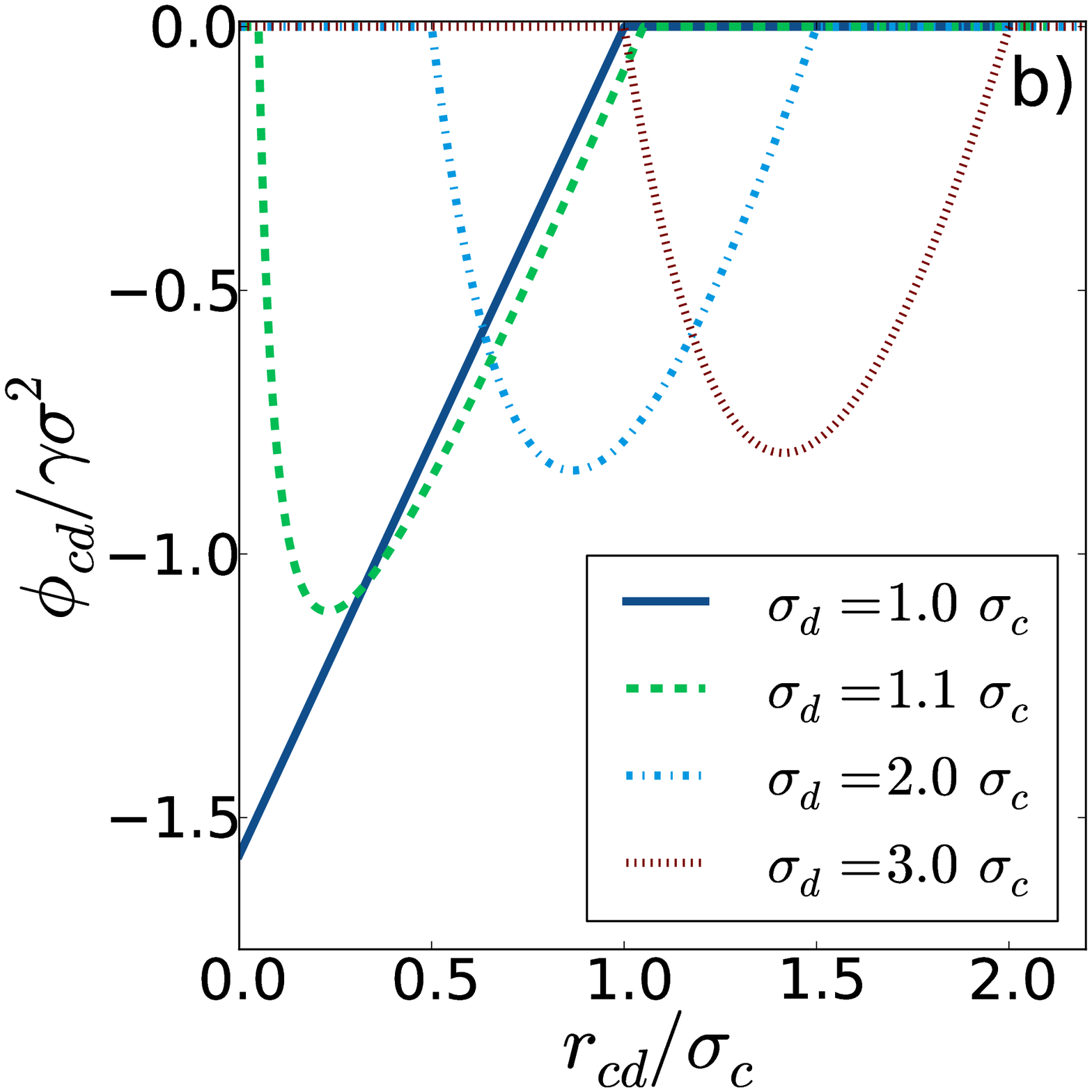}
\caption{Pair interactions for the binary mixture of colloids and droplets. a) Comparison between the Asakura-Oosawa-Yukawa (AOY)  and the Square-Well-Yukawa (SWY) potentials with $q=0.1$, $\kappa \sigma_{c}$=10, $ \beta \epsilon_{Y}$=24.8, $\beta \epsilon_{AO}=2.1$, $\beta \epsilon_{SW}=9$. These parameters were chosen in order to 
obtain $|\rm min(\beta \phi_{\rm cc})|=\rm max(\beta \phi_{\rm cc}) =9$.
b) Colloid-droplet potential $\phi_{cd}/\gamma \sigma^2$ scaled by the droplet-solvent interfacial tension $\gamma$ for $\sigma_{d}/\sigma_{c}=3, 2, 1.1$, and 1 (from right to left).
 }
\label{fig:pot}
\end{figure}

\subsection{Definition of the pair interactions}
\label{s:int}
We study a binary mixture of $N_{c}$  colloidal particles with hard-sphere diameter $\sigma_{c}$ and $N_{d}$ droplets of diameter $\sigma_{d}$. 
The total interaction energy $U$ is the sum of colloid-colloid, droplet-droplet and colloid-droplet interactions, 
\begin{eqnarray}
U&=&\sum_{i<j}^{N_{c}} \phi_{cc}(|\vec r_{i}-\vec r_{j}| )+\sum_{i<j}^{N_{d}}  \phi_{dd}(|\vec R_{i}-\vec R_{j}| ) \nonumber \\
 && + \indent \sum_{i}^{N_{c}} \sum_{j}^{N_{d}}  \phi_{cd}(|\vec r_{i}-\vec R_{j}|), 
\end{eqnarray}
where $\vec{r}_i$ is the center-of-mass position of colloid $i$, $\vec R_{j}$ is the  center-of-mass position of droplet $j$, $\phi_{cc}$ is the colloid-colloid pair interaction, $\phi_{cd}$ is  the colloid-droplet pair interaction, and $\phi_{dd}$ is the droplet-droplet pair interaction.

We consider two different types of colloid-colloid interactions. 
The first is the sum of the short-ranged attractive Asakura-Oosawa potential $U_{\rm AO}(r)$ and the long-ranged repulsive Yukawa potential $U_{\rm Y}(r)$, i.e.
\begin{equation} 
 \phi_{\rm cc}(r)=\left \{ 
\begin{array}{ll}
 \infty  &  r<  \sigma_c\\
U_{\rm Y}(r) +  U_{\rm AO}(r)
    &  \text{otherwise,} \\
\end{array} \right . 
\label{eq:AOY}
\end{equation}
with $\beta=1/k_{B} T$, where $k_{B}$ is the Boltzmann constant and
$T$ the temperature.
Here the Yukawa potential is defined by
\begin{equation}
U_{\rm Y}(r) =  \epsilon_{\rm Y}  \sigma _{c} \frac {e^{- \kappa
               (r -\sigma_c)}}{r} ,
                \label{YU}
\end{equation}
and the Asakura-Oosawa potential~\cite{Asakura1954,Vrij1976} is

\begin{equation}
 U_{\rm AO}(r)=\left \{ 
\begin{array}{ll}
 -\epsilon_{\rm AO} f(r)   & \, \, \sigma_{c}< r< \sigma_{c}(1+q)  \\
 0 &  \text{ otherwise,} 
 \end{array} \right . 
\label{te:ao}
\end{equation}
with $f(r)= \left [ 1- \frac{3 r}{2
(1+q)\sigma_c} + \frac{r^3}{2(1+q)^3\sigma_c^3}\right ] $. 
The parameter $\epsilon_{\rm AO}$ controls the strength of attraction, while $\epsilon_{\rm Y}$ controls the strength of the repulsion. The range of the interactions are controlled by the parameter $q$ for the AO potential and by $\kappa$ for the Yukawa interaction. We refer to the combined interaction (\ref{eq:AOY}) as the  Asakura-Oosawa-Yukawa (AOY) potential. 
Although both the AO and Yukawa potentials have very specific physical interpretations, we use them in this paper merely as  generic models for a steep short-range attraction and a long-range repulsion, respectively. 
The AO potential describes the depletion attraction between colloidal particles due to the presence of non-adsorbing polymers with radius of gyration $\sigma_{c} q/2$. In the limit of small $q$ Eq.(\ref{te:ao}) is exact and $f(r)$ is related to the free volume gained by the polymers when two colloids are close to each other~\cite{Dijkstra1999a}.  
The Yukawa potential describes the interaction between two charged particles screened by a medium with inverse Debye length $\kappa$. 
As shown in  Fig.~\ref{fig:pot}a), for a typical set of parameters (justified below) that we used in the simulations, the potential $ \phi_{\rm cc}(r)$ defined by Eq.~(\ref{eq:AOY}) has a maximum at distance $r_{\rm peak}$. 
In order to investigate the influence of the shape of the attractive part of the potential we also consider a modified version of  (\ref{eq:AOY}) using a square well potential for distances smaller than $r_{\rm peak}$.
This is 
\begin{equation} 
 \phi_{\rm cc}(r)=\left \{ 
\begin{array}{ll}
 \infty  &  r<  \sigma_c\\
-  \epsilon_{SW}
    &  \sigma_c<r<  r_{\rm peak} \\
  U_{Y}(r) +  U_{\rm AO}(r)
    & \text{otherwise,} 
    \end{array} \right . 
\label{eq:SWY}
\end{equation}
which we refer to as the Square-Well-Yukawa (SWY) potential. It is shown in Fig.~\ref{fig:pot} with a dashed line.  This potential is very similar to that used by~\citet{Mani:2010p4140} in their pioneering  study of the stability of colloidal shells. 
The parameter space of the interaction is arbitrarily restricted to potentials with the shape shown in Fig.~\ref{fig:pot}, i.e. with $\rm max(  \phi_{\rm cc}) = |\rm min(\phi_{\rm cc})|$, and hence $\phi_{\rm cc}(r_{\rm peak}) = - \phi_{\rm cc}(\sigma_{c})$. The height of the repulsive barrier, $\phi_{\rm cc}(r_{\rm peak})  - \phi_{\rm cc}(\infty)$, is half the depth of the attractive well, $\phi_{\rm cc}(r_{\rm peak})  - \phi_{\rm cc}(\sigma_{c})$.

The droplet-droplet interaction is taken to be hard-core repulsion 
\begin{equation} 
 \phi_{\rm dd}(r)=\left \{ 
\begin{array}{ll}
 \infty  &  r<  \sigma_d+\sigma_{c}\\
  0 & \text{otherwise,} 
\end{array} \right . 
\label{eq:dd}
\end{equation}
with an effective hard-core diameter $\sigma_d+\sigma_{c}$ that is larger than the bare droplet diameter $\sigma_{d}$. Using the effective diameter ensures that the surface-surface distance between any two droplets is always larger than one colloid diameter. In this way two droplets can never bind together due to a shared colloid.

The colloid-droplet interaction is aimed at modeling the Pickering effect. The loss of interfacial  energy~\cite{Pieranski1980} when a particle is trapped at the surface of the droplet is $\gamma S$, with $S$ the droplet surface that is covered by the colloid, and $ \gamma$ the droplet-solvent interfacial tension. 
The relation is valid when the interfacial tension between the colloid and the droplet is the same as that between the colloid and the solvent.
The surface $S$ has different expressions depending on the size of the droplet and the colloid-droplet separation. 
If the diameter of the droplets is larger than the diameter of the colloidal particles, i.e. for $\sigma_d > \sigma_{c}$, the colloid-droplet energy is
\begin{equation} 
\phi_{\rm cd}(r)= \left \{ 
\begin{array}{ll}
-     \gamma  \pi \sigma_{d} h &  \frac{\sigma_d-\sigma_{c}}{2}<r<  \frac{\sigma_d+\sigma_{c} }{2}\\
    0 & \mbox{otherwise,}
\end{array} \right . 
\label{eq:DC}
\end{equation}
with $h=(\sigma_{c}/2-\sigma_{d}/2+r)(\sigma_{c}/2+\sigma_{d}/2-r)/(2r) $  the height of the spherical cap that results from the colloid-droplet intersection.
On the other hand, when the diameter of the droplets is smaller than the diameter of the colloidal particles, i.e. $\sigma_d < \sigma_{c}$, we assume
\begin{equation} 
\phi_{\rm cd}(r)= \left \{ 
\begin{array}{ll}
-    \gamma  \pi  \sigma_{d}^{2} &  r< \frac{\sigma_c-\sigma_{d}}{2}  \\
-     \gamma  \pi \sigma_{d} h &  \frac{\sigma_c-\sigma_{d}}{2}<r<  \frac{\sigma_c+\sigma_{d} }{2}\\
    0 & \mbox{otherwise.}
\end{array} \right .
\label{eq:DC1}
\end{equation}
Within this model we neglect the influence of the particle on the oil-water interfacial curvature. The contact angle changes upon changing the position of the particle with respect to the oil-water interface, while in reality the contact angle remains constant and the curvature of the oil-water interface changes, such that the droplet becomes non-spherical, which is beyond our model.

\subsection{Simulation method}
\label{s:sim}
 
In our model we neglect the coalescence of the droplets and the hydrodynamic interactions due to the solvent. Even with these simplifications, the relevant time scale is not easily reachable in  standard Brownian Dynamics (BD) simulations. Hence, we carry out Metropolis Monte Carlo (MC) simulations, that for small displacement steps can reproduce the correct dynamics~\cite{Sanz:2010p3489} and are much more efficient to run than BD simulations. The evolution of the system is therefore described by the number of MC sweeps per particle. 
For the colloidal particles~\cite{Sanz:2010p3489}  the MC maximum trial displacement $d_{c}$ is set to $d_{c}=0.01 \sigma_{c}$. 

We define a bond between two colloidal particles when their distance is smaller than $r_{\rm peak}$. 
A cluster is defined as a set of  colloidal particles connected by a network of bonds.
The translational diffusion of the clusters is achieved by  translational MC cluster moves~\cite{WU:1992p4146} with maximum linear displacement $d^{t}_{\rm cls}=d_{c}/\sqrt[6]{N}$ with $N$ the number of particles in the cluster. This approximates the hydrodynamic slowing down of a spherical cluster that satisfies the Stokes-Einstein equation for the diffusion constant $D=\frac{k_{B}T}{3 \pi \eta \sigma_{\rm cls}}$, with $\eta$ the viscosity of the solvent and  $\sigma_{\rm cls}$ the diameter of the sphere that approximate the shape of the cluster. Here, we assume that  $\sigma_{\rm cls}  \sim \sqrt[3]{N}$.
Additionally, we mimic the cluster rotational diffusion via  rotational MC moves, in which clusters are rotated around a random axis with a maximum angle $d^{r}_{\rm cls}= 0.01 \sigma_{c}/\sigma_{\rm cls}$.
In a cluster move~\cite{WU:1992p4146,Frenkel:2002p2093}, all particles belonging to a cluster are translated or rotated by the same amount. In order to satisfy the condition of detailed balance all cluster moves that lead to two clusters merging or a cluster and a single particle merging are rejected.

The droplets move according to the MC scheme with a maximum displacement $d_{d}=d_{c} \sqrt{\sigma_{c}/\sigma_{d}}$. 
The evaporation dynamics is introduced by forcing the droplets diameter $\sigma_{d}$ to shrink at a fixed rate. The rate is chosen so that the droplets vanish half-way though the simulation  ($5 \times 10^{5}$ sweeps). This leaves another $5 \times 10^{5}$ sweeps to investigate the stability of the clusters against thermal fluctuations. Here the timescales are chosen for practical reasons. In the experiments stability can be relevant on the time scales of years, while the clusters experiments last typically tens of minutes. Hence, our simulation do not address the true long-time behavior of the system. 

We restrict ourselves to symmetric potentials with $\rm max(\beta \phi_{\rm cc}) = | \rm min(\beta \phi_{\rm cc})|$. 
In particular, we investigate the case ${\rm max}(\beta \phi_{\rm cc}) = 9$ as shown in Fig.~\ref{fig:pot}  ($q=0.1$,  $\kappa \sigma_{c}=10$ and $r_{c}=2.5 \, \sigma_{c} $). 
The height of the repulsive barrier is $9 k_{B} T$, a value big enough  to hinder spontaneous clustering, while the depth of the attractive well  is $18 k_{B} T$, so that in practice a particle cannot escape by thermal fluctuations. 
For these parameters we find that the maximum of the potential is at $r_{\rm peak}=1.0845 \sigma_{c}$.
For each parameter set we run eight independent MC simulations with $N_{c}=500$ colloidal particles with packing
 fraction $\eta_{c}=0.0034$, and 0.01. The droplets packing fraction is fixed at $\eta_{d}= 0.1$. Simulations have been performed for initial droplet sizes $\sigma_{d}(0)=2,4,6$, and $8 \sigma_{c}$. Each simulation consisted of $10^{6}$ MC sweeps. In every sweep all particles are attempted to be moved on average once. 
The droplets and the particles are initialized randomly in the cubic simulation box, with the constraints that all colloids are outside of the droplets and that the minimum distance between colloidal particles is  larger than $r_{\rm peak}$. Hence, we start in a state without clusters. 
We characterize the cluster structure by the bond-number $n_{b}$, corresponding to the number of pairwise bonds in a cluster. The number of bonds is also an estimate of the total energy of the cluster; a higher number of bonds corresponds to a greater attractive energy.

\subsection{Experimental methods}
\label{s:exp}
\subsubsection{Cluster preparation}
The colloidal particles are positively charged and narrowly dispersed polystyrene spheres with 154 nm diameter. A detailed description of the preparation of the constituent particles and their assembly into clusters is given in Ref.~\cite{Wagner:2008p844}. In particular, in these experiments, the particles were added both via the water and via the oil phase. 
For the present studies the clusters were prepared in a slightly modified fashion. Briefly, 53 mg polystyrene particles suspended in 3 ml of toluene and 73 $\mu$l dodecane (to suppress Ostwald ripening) were emulsified with 27 ml of an aqueous solution of Pluronic$\textsuperscript{\textregistered}$ F-68 (1 wt.\%) using an ultrasonic homogenizer (Sonoplus HD 3200, Bandelin). Evaporation of the dispersed toluene phase under reduced pressure (50 mbar, 40 $^\circ$C) initiated the assembly of the particles into clusters.

\subsubsection{Electron microscopy of droplet and assembly morphology}

The emulsion droplets bearing polystyrene particles at their surface were examined on a cryogenic field emission scanning electron microscope (Ultra Plus, Zeiss). Specimen preparation was accomplished by sandwiching 4  $\mu$l of the emulsion in between two aluminum platelets (3 mm x 0.5 mm, 0.15/0.15 mm, Engineering Office M. Wohlwend). The carrier assembly was plunged into a high-pressure-freezing machine (EM HPM100, Leica) and was vitrified at 2000 bar within 20 ms. This helped sealing the sandwich so that nucleation of ice crystals and specimen damage were suppressed. In a cryo preparation chamber (EM MED020 FF, Leica) the sample was freeze-fractured, lightly etched for 60 s at $-112 \ ^\circ$C, and sputtered with platinum in an amount equivalent to a 4 nm thick coating. The specimen was transferred by a cryo shuttle (EM VCT 100, Leica) to the cold stage of the microscope. Micrographs were recorded digitally at a temperature of $-160 \ ^\circ$C, with an aperture of 10  $\mu$m and a voltage of 1.0 kV. 
The morphologies of the colloidal assemblies were analyzed by field emission scanning electron microscopy (FESEM) on a Zeiss LEO 1530 Gemini microscope equipped with a field emission cathode operating at 3 kV. A minute amount of the cluster suspension (10$^{-5}$ wt.\%) was placed onto a silicon wafer (CrysTec) and dried under ambient conditions. The specimen was coated with a platinum layer of 1.3 nm thickness using a sputter coater (Cressington 208HR) to make the specimen conductive. 

\section{Results}

\subsection{Dynamics of cluster formation}
\label{s:fluid}

\begin{figure*}
\includegraphics[width=16cm]{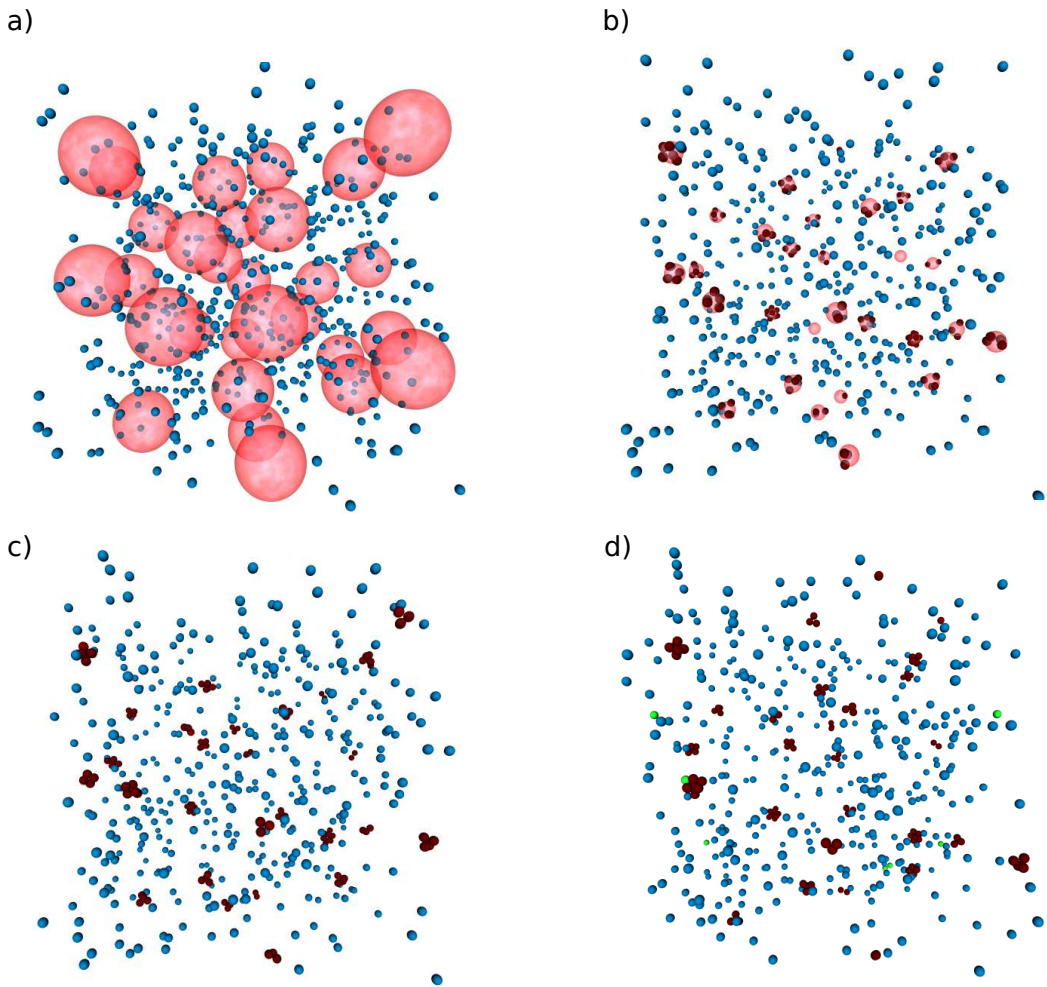}
\caption{Snapshots of the binary mixture of colloids and droplets at colloid packing fraction $\eta_{c}=0.0034$ and SWY potential. Shown are results at different stages of the simulation. a) Initial configuration, b) after $3.6\times 10^{5}$,  c) after $5 \times 10^{5}$,  d) after $10^{6}$ MC sweeps. Droplets are shown as pink spheres, single colloidal particles are depicted as blue spheres, green is used for  particles belonging to spontaneously formed clusters. Colloidal particles trapped at the surface of a droplet or in a droplet-induced cluster are shown in red. A movie of the simulation is available \href{http://www.youtube.com/watch?v=R9wlarGvJmQ&context=C2a833ADOEgsToPDskLFofSSFhvugIcYhueO99zW}{here.}}
\label{fig:snp}
\end{figure*}

Figure~\ref{fig:snp} shows simulation snapshots at four different stages of the simulation.  
The initial configuration of the simulation (Fig.~\ref{fig:snp}a) is a binary mixture of non-overlapping spheres. The large (pink) spheres represent the droplets, while the small (blue) spheres represent single colloidal particles.  After $3.6\times 10^{5}$ MC sweeps (Fig.~\ref{fig:snp}b) particles are trapped at the surface of a droplet (red). 
Figure~\ref{fig:snp}c) shows the configuration after $5\times 10^{5}$ sweeps, just after the droplets have vanished completely. Figure~\ref{fig:snp}d) shows the configuration at the end of the simulation after $10^{6}$ sweeps. All clusters that are formed formed due to the droplets are still present in the system demonstrating the stability of the clusters against thermal fluctuations. Few doublets (green) have also formed spontaneously.   \href{http://www.youtube.com/watch?v=R9wlarGvJmQ&context=C2a833ADOEgsToPDskLFofSSFhvugIcYhueO99zW}{See also a movie of the simulation here.}
As a check we simulated the structure of the pure colloidal fluid without droplets.  
We find that the structure of the single component fluid is largely composed of single particles with only 1\% of particles belonging to doublets.  Larger clusters are not formed during the span of our simulations. 
This can be explained easily by the following argument. The minimum distance between colloidal particles is at the beginning of the simulation larger than $r_{\rm peak}$. In order to form a bond, the colloidal particles have to overcome the repulsive energy barrier ($9 k_{B}T$) of the colloid-colloid interaction. The probability to thermally overcome this barrier is very small, and in order to form larger clusters the particles have to be forced beyond the repulsive barrier. Hence, in the restricted time interval that is accessible in the simulations, the colloidal fluid is (quasi-)stable. However, for longer times further clustering might occur (see discussion in the Sec.~\ref{s:sim}).

\begin{figure}
\includegraphics[width=8cm]{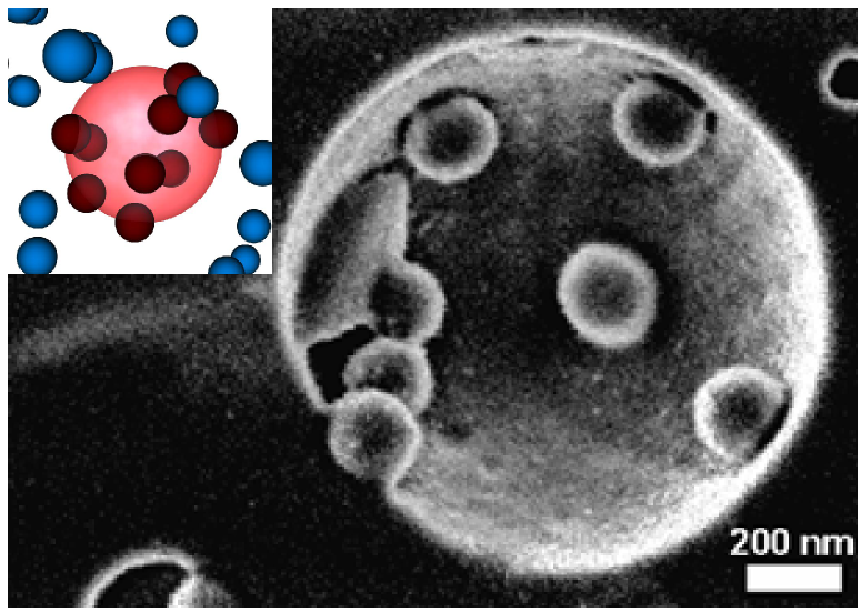}
\caption{
Colloidal particles trapped at the surface of an emulsion droplet obtained with a Cryo-FESEM micrograph of a toluene-in-water emulsion stabilized by crosslinked polystyrene particles. Inset: Simulation snapshot of a single droplet and colloidal particles trapped at its surface obtained after $3 \times10^{5}$ MC sweeps.}
\label{fig:cry}
\end{figure}

In the binary mixture, a clustering mechanism is provided by the shrinking droplets.
The formation of small clusters implies that only a limited number of particles are bound onto the droplets. 
 The small size of our particle makes them ideal for the
  self-assembly of small clusters that are well suited as building
  blocks for subsequent self-assembly, but prevented us to follow the assembly process in experiments in real
  space like~\citet{Manoharan:2003p937}.
Therefore, we studied their distribution in the experimental emulsions with cryo-FESEM. The micrographs (Fig.~\ref{fig:cry}) indicate a random distribution of the positions of the particles at the droplet surface. The void in the center presents the imprint of the frozen dispersed phase, i.e. a single micron-sized toluene droplet. The polystyrene particles left in the cavity after sublimation of toluene are randomly distributed at the former droplet surface. Because toluene is a good solvent for polystyrene, the particles are significantly swollen at the droplet interface as expressed by a larger diameter. This agrees with the fact that the micrograph indicates that the polystyrene particles  prefer the dispersed toluene phase.
The position distribution of the particles on the droplet surface in simulations (Fig.~\ref{fig:snp}b and inset Fig.~\ref{fig:cry}) is in good agreement with the experiment and indicates that the particles can freely diffuse on the surface of the droplets. 

In simulations, we consider the case that the particle surface has no preference of whether it wets oil or water. Hence the contact angle at a (planar) oil-water interface is 90$^{\circ}$ because of the assumption that the colloid-solvent interfacial tension is equal to the colloid-droplet interfacial tension -- see the discussion above Eq.(\ref{eq:DC}) . 
Despite the difference in contact angle between simulation and experiments we do not expect the contact angle to affect the final results~\cite{Lauga2004}.

The liquid structure can be further characterized in computer simulations. We calculated the colloid-droplet radial distribution functions, $g_{cd}(r)$, and the colloid-colloid radial distribution functions, $g_{cc}(r)$, at different stages of the simulations. Since the droplet diameter $\sigma_{d}$ changes continuously  during the simulation the resulting transient structures captured by the distribution functions are not at equilibrium. 
\begin{figure}
\begin{tabular}{cc}
\includegraphics[width=4.5cm]{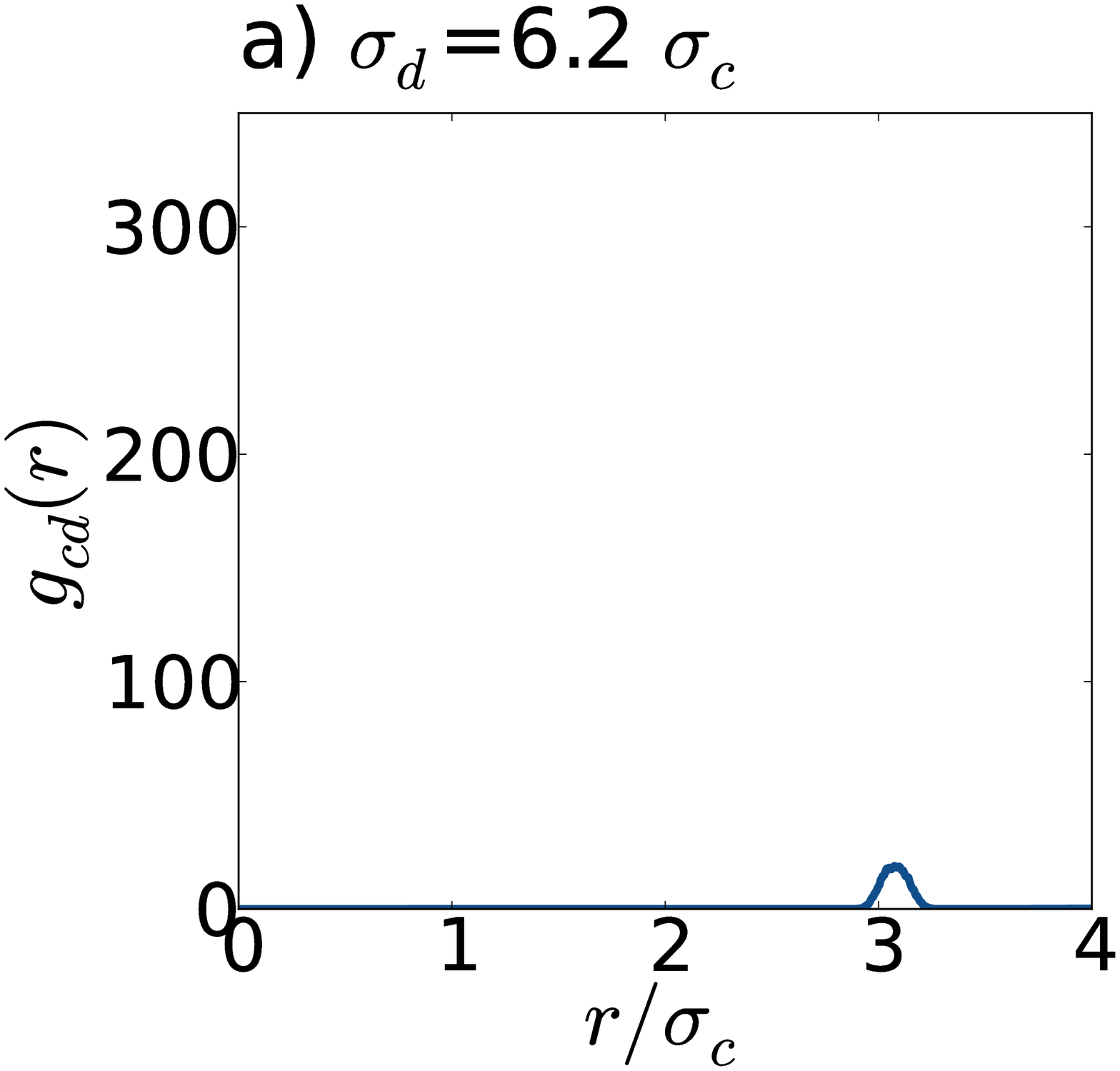} &
\includegraphics[width=4.5cm]{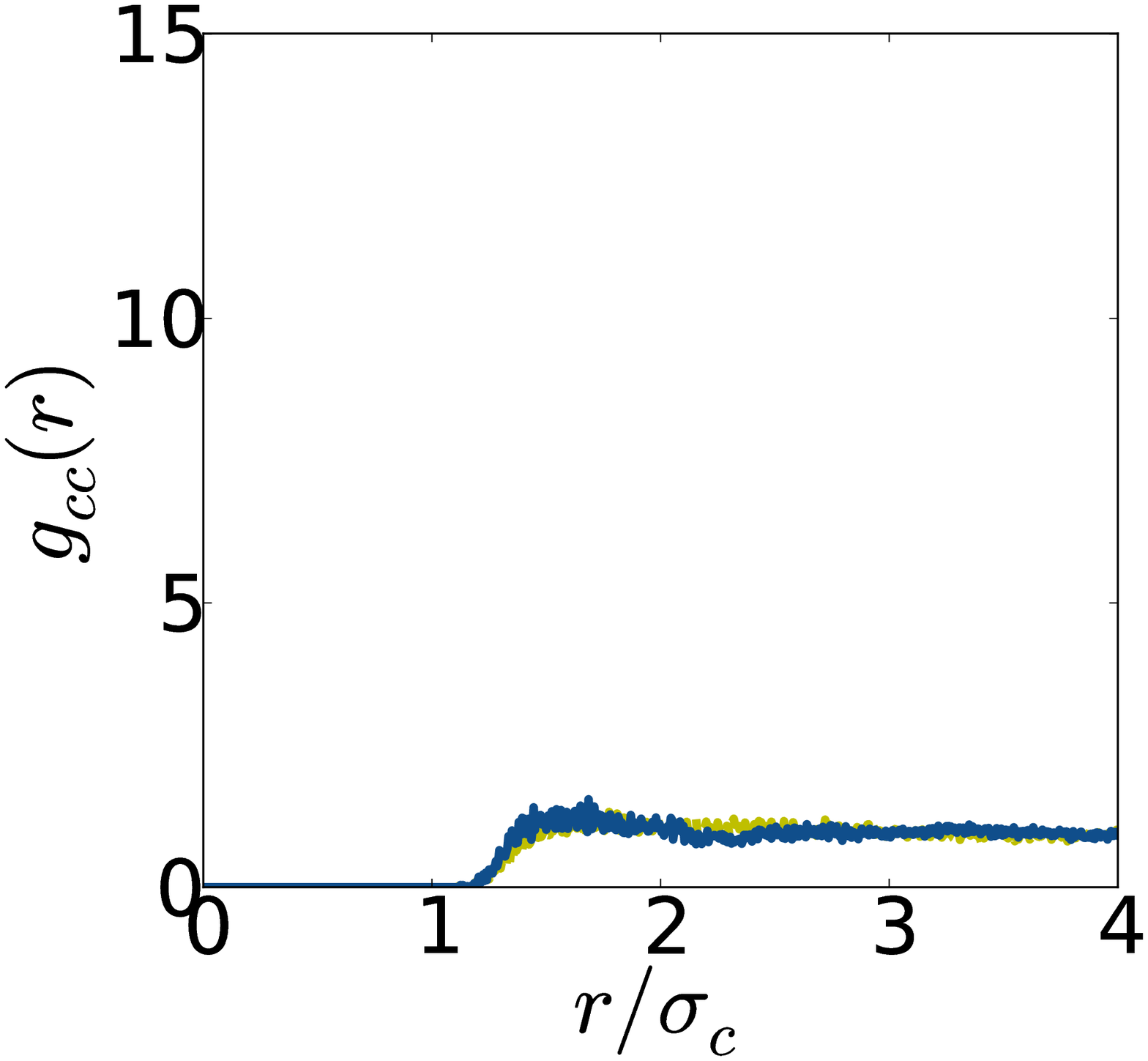}\\
\includegraphics[width=4.5cm]{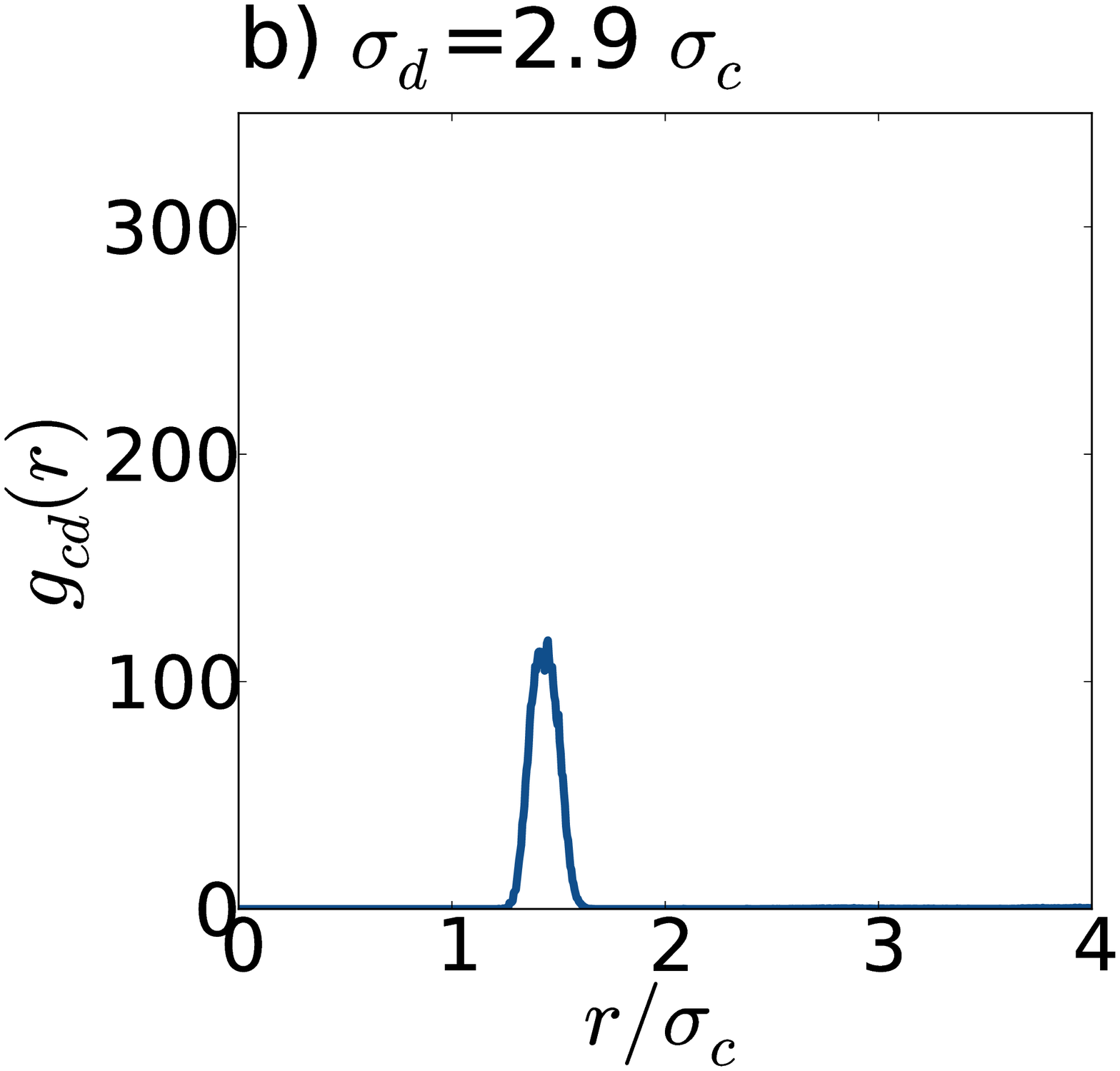}&
\includegraphics[width=4.5cm]{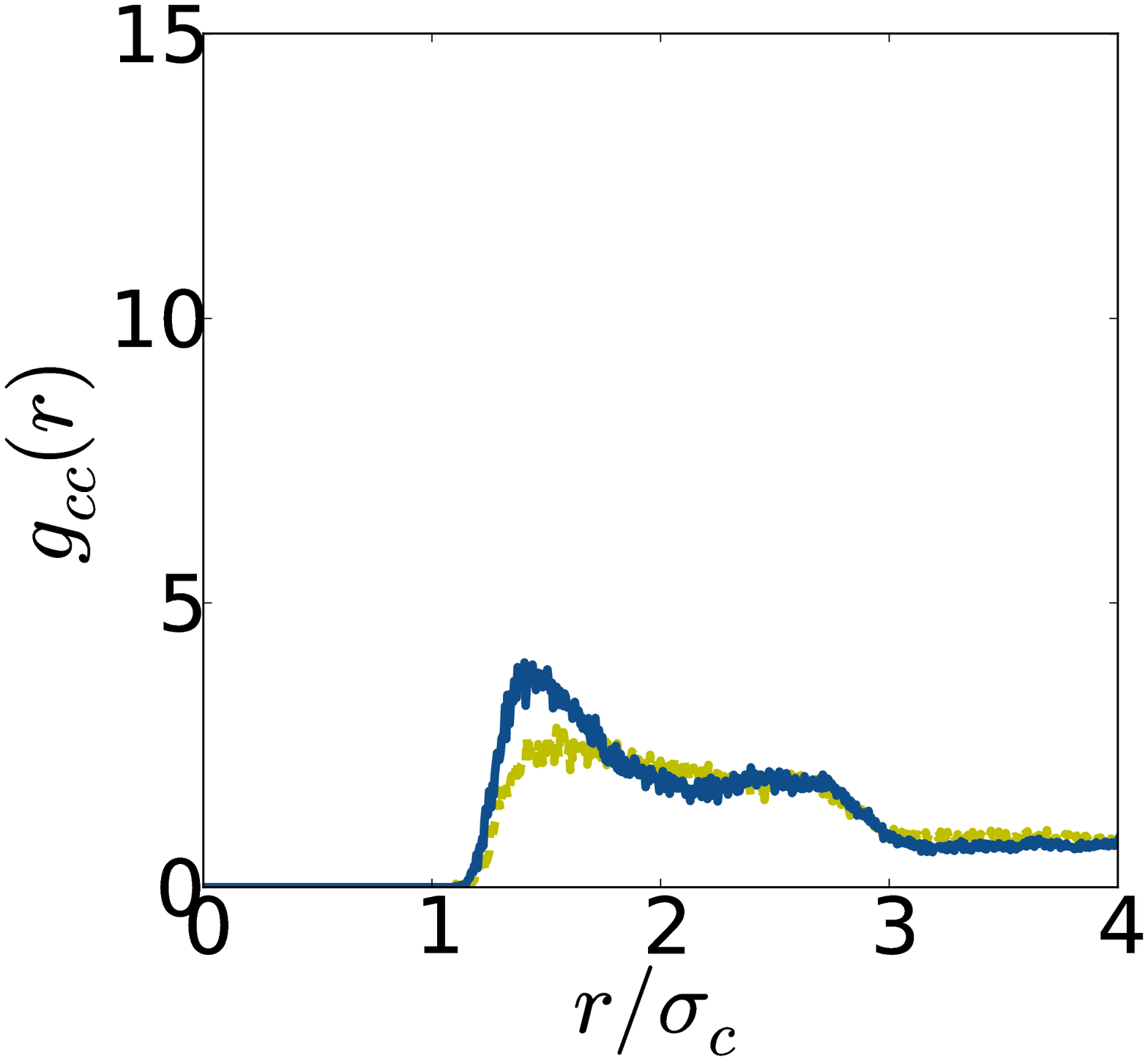}\\
\includegraphics[width=4.5cm]{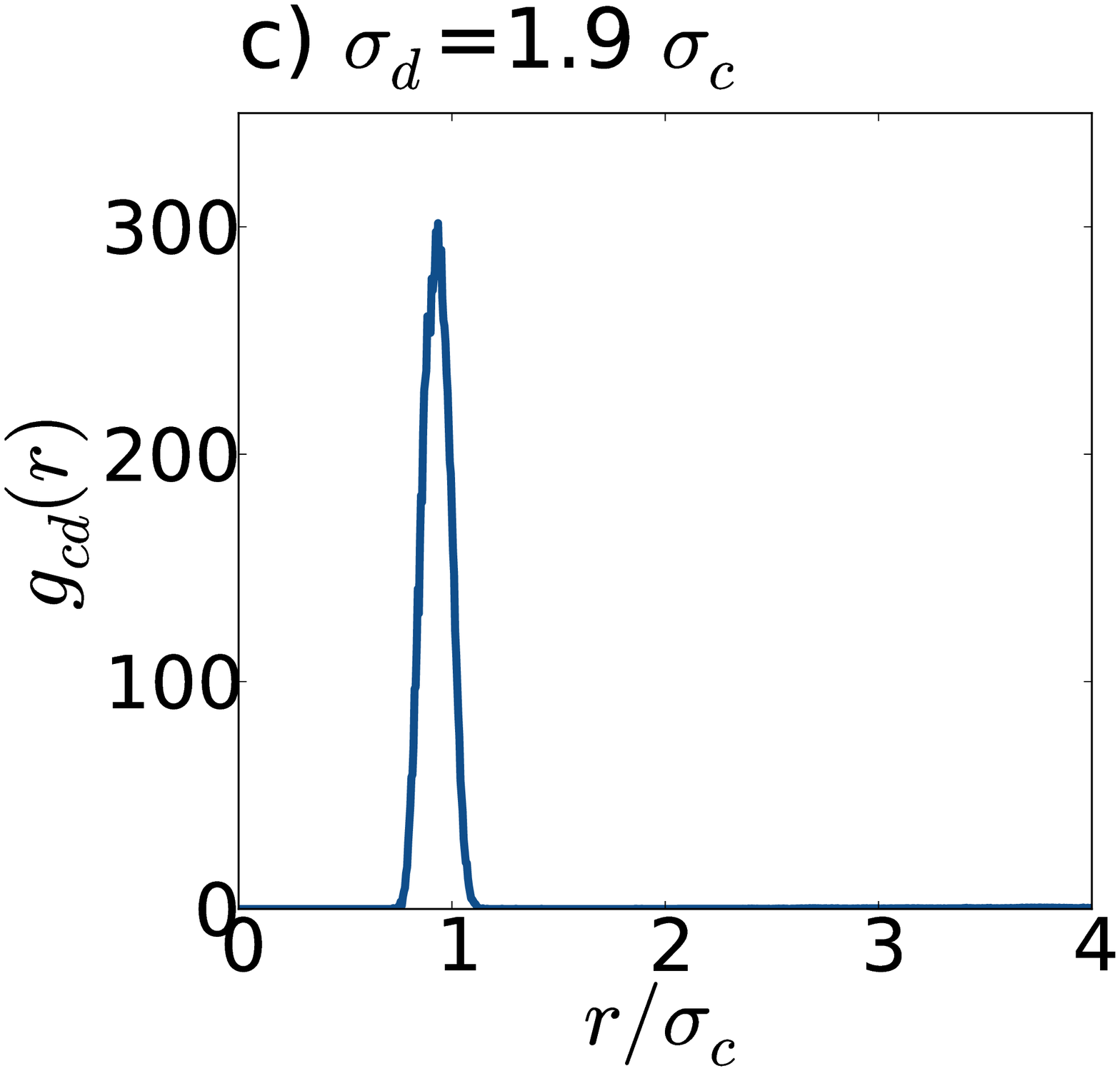}&
\includegraphics[width=4.5cm]{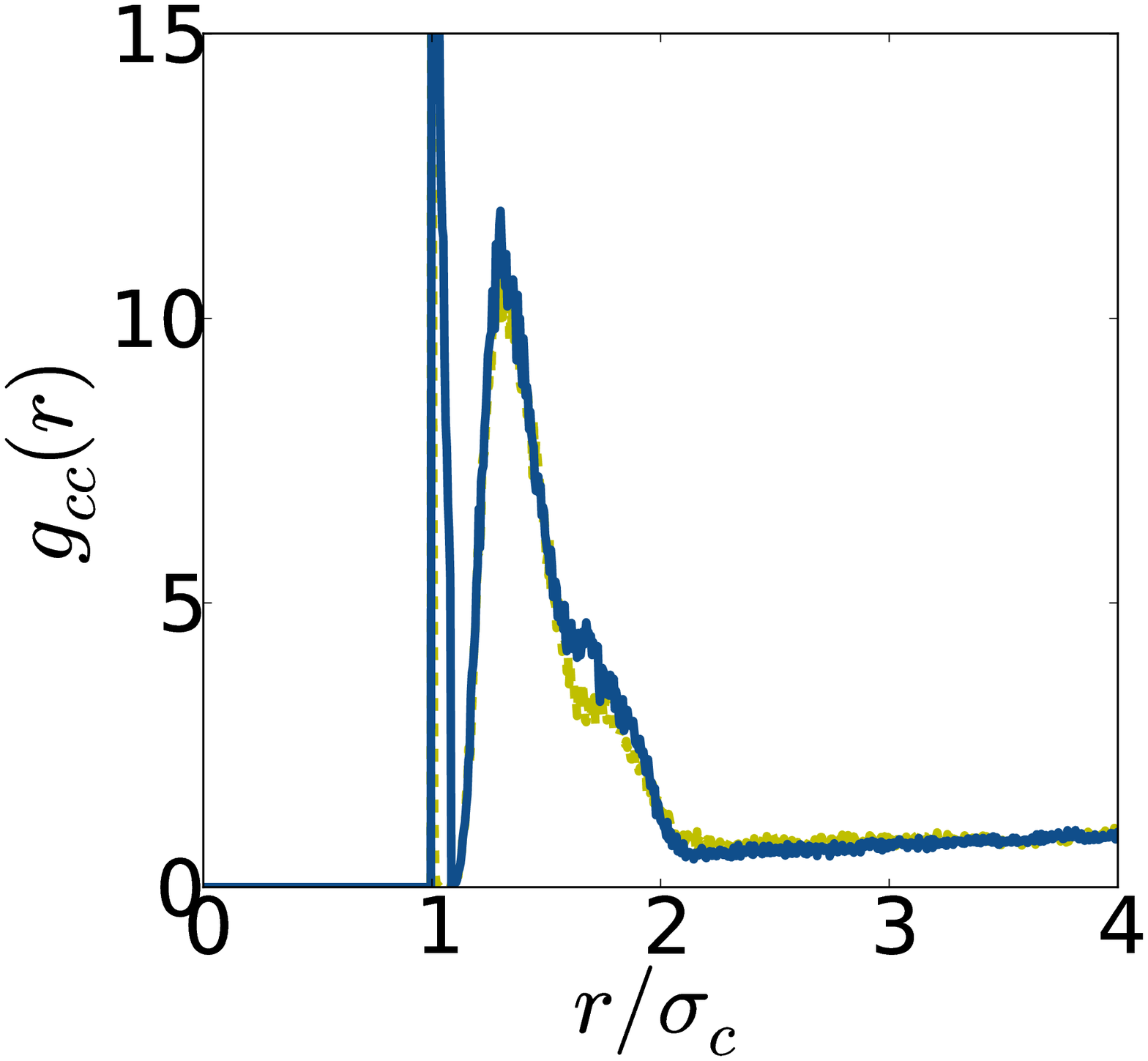}\\
\includegraphics[width=4.5cm]{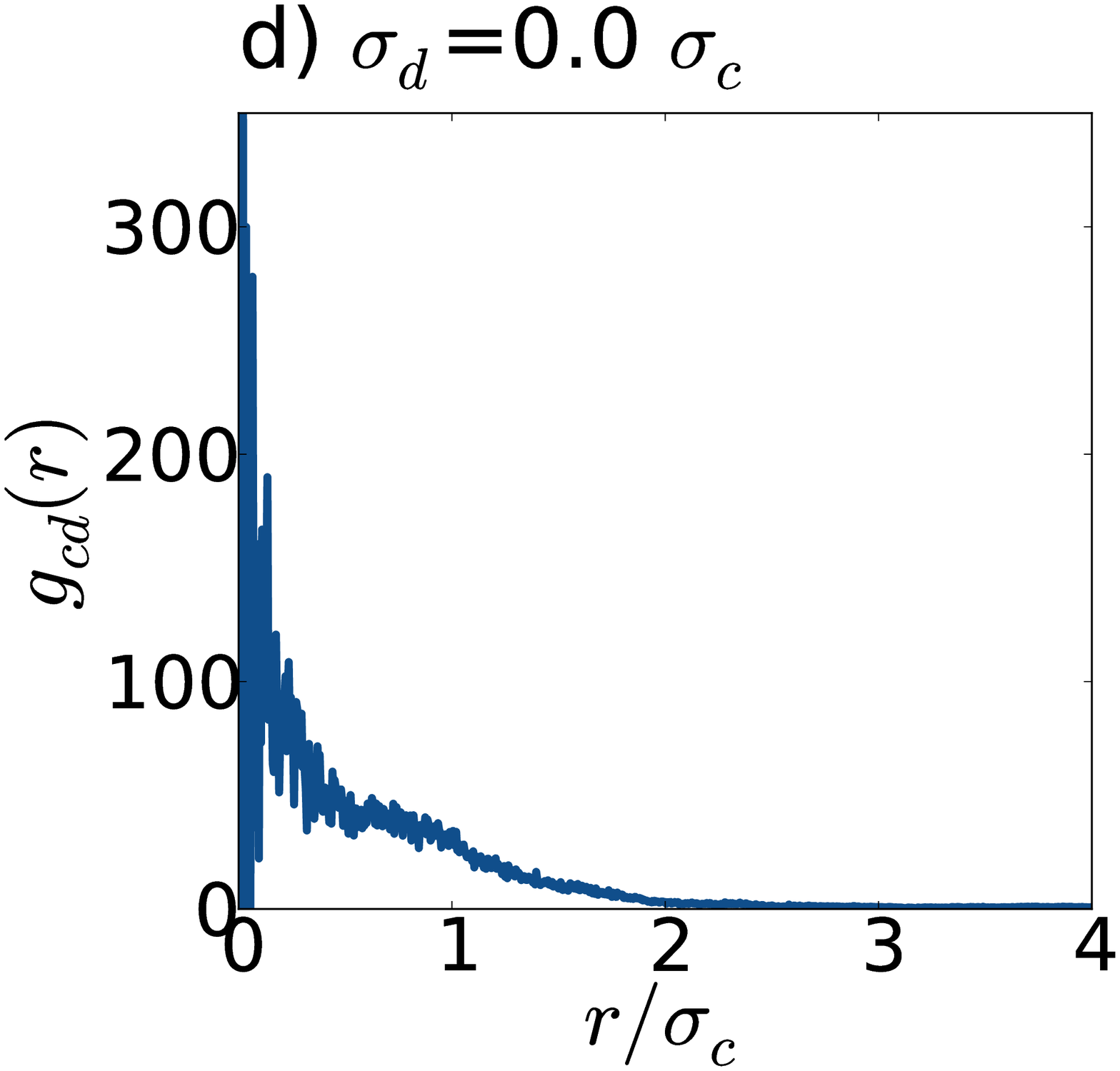}&
\includegraphics[width=4.5cm]{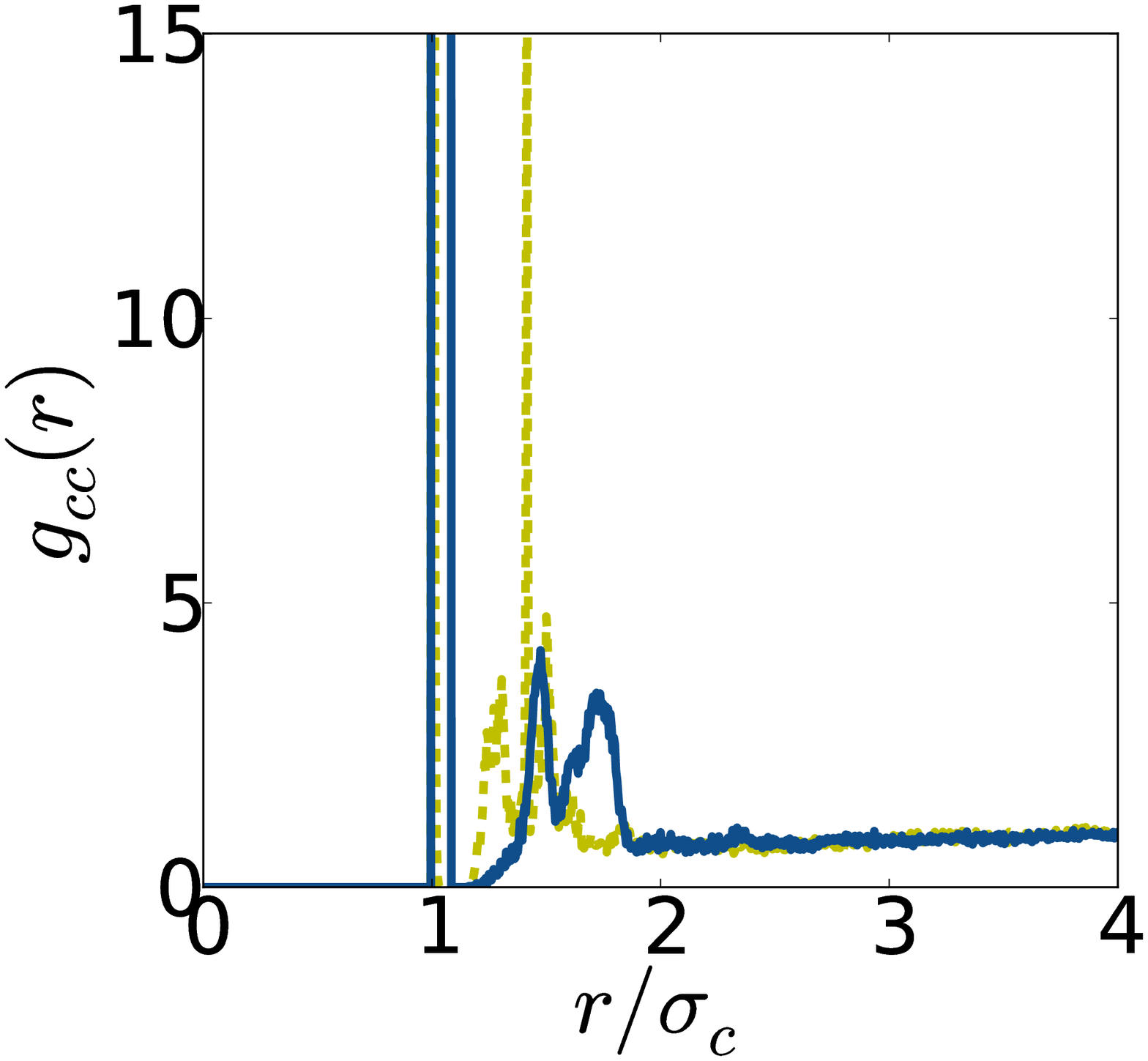}\\
\end{tabular}
\caption{Colloid-droplet (left column) and colloid-colloid (right column) radial distribution functions, $g_{cd}(r)$ and $g_{cc}(r)$, respectively, as a function of the scaled distance $r/\sigma_{c}$ at colloid packing fraction $\eta_{c}=0.0034$. We plot  $g_{cc}(r)$  for the SWY (blue full line) and the AOY (yellow dashed line) potentials. Shown are results  at different stages of the computer simulation: a)  after $10^{5}$, b) after $3 \times 10^{5}$, c) after $3.6\times 10^{5}$, d) after $5 \times 10^{5}$ MC sweeps.}
\label{fig:gr}
\end{figure}
%%%%%%%%%%%%%
Fig.~\ref{fig:gr} shows  $g_{cd}(r)$ and $g_{cc}(r)$ at different stages. For the colloid-colloid correlation function we plot both the results of the SWY (blue full line) and AOY potentials (yellow dashed line).
In particular, Fig.~\ref{fig:gr}a) shows the correlation functions after $10^{5}$ MC sweeps. 
The colloid-droplet radial distribution function $g_{cd}(r)$ has a small peak at $r \simeq 3.1 \ \sigma_{c}$ corresponding to the instantaneous droplet radius $\sigma_{d}(t)/2$. The peak is due to colloidal particles trapped at the droplet surface. At the same time the colloid-colloid radial distribution functions $g_{cc}(r)$ are apparently flat outside of the core region. In fact, the radial distribution function is decaying in a way consistent with the Boltzmann factor $g(r)=e^{-\beta \phi_{cc}(r)}$ as expected for an equilibrium low density gas.   
Figure~\ref{fig:gr}b) shows the correlation functions after $3 \times 10^{5}$ MC sweeps. The cross pair correlation function $g_{cd}(r)$ shows that the droplets have shrunk to a radius $\sigma_{d}(t)/2\simeq 1.45 \ \sigma_{c}$, while $g_{cc}(r)$ has developed structure at intermediate distances. These results can be explained by particles trapped at the surface of the droplets and interacting with each other via the long-ranged colloid-colloid repulsion. 
After $3.6\times 10^{5}$ MC sweeps (Fig.~\ref{fig:gr}c) the droplet radius has become smaller than  $\sigma_{c}$. The strong peak at $r = \sigma_{c}$ indicates that a large number of bonds between colloidal particles have formed for both the AOY and SWY potentials. Finally, after $5 \times 10^{5}$ MC sweeps (Fig.~\ref{fig:gr}d)  a dramatic change of $g_{cd}(r)$ is observed, which is due to droplets having a diameter  $\sigma_{d}(t) =0$, and diffusing freely. We show the left panel of Fig.~\ref{fig:gr}d) for completeness, but stress that it does not correspond to any physical situation. On the other hand, $g_{cc}(r)$ shows that strong peaks have formed also at distances larger than $\sigma_{c}$, indicating the  presence of small clusters. 
The time evolution of $g_{cc}(r)$ for the two potentials is very
similar. Most notably the final configurations differ due to different final cluster structures. 

\subsection{Cluster structure and size distribution}
\label{s:cls}
\begin{figure}
\includegraphics[width=9cm]{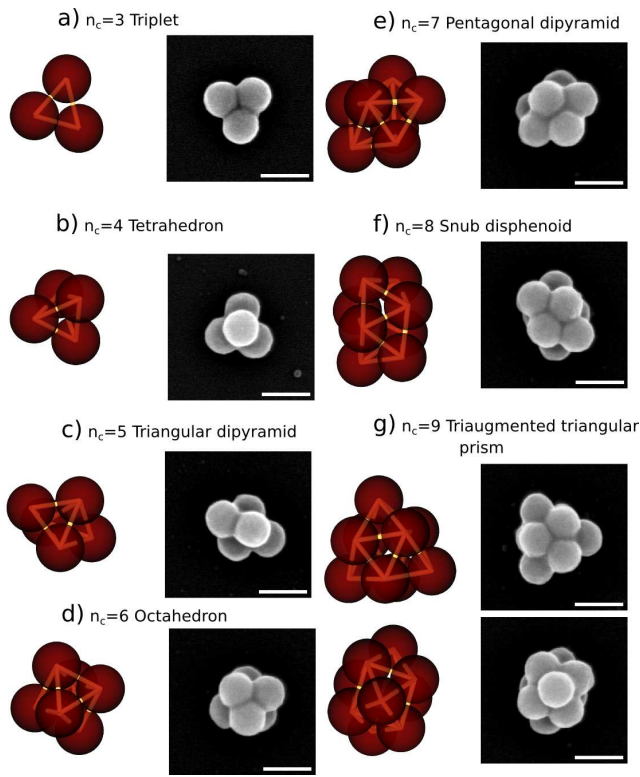}
\caption{Cluster structures found in simulations (left)  for SWY potential, $\sigma_{d}(0)=8 \sigma_{c}$, $\eta_{c}=0.0034$ and micrographs from FESEM (right). The scalebars indicate 200nm. The wireframe in the simulation structures connects the particles centers in order to visualize the geometric arrangement.}
\label{fig:snap1}
\end{figure}

\begin{figure}
\includegraphics[width=9cm]{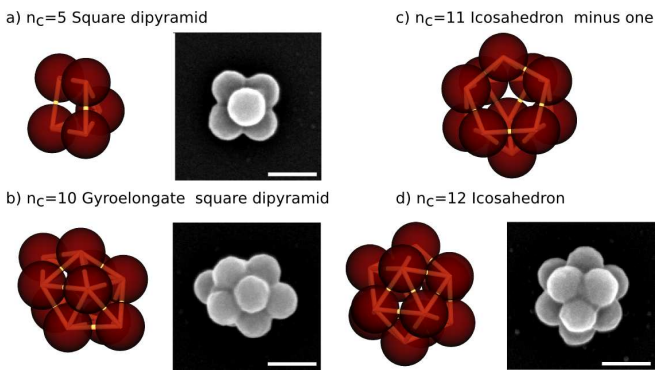}
\caption{Same as Fig.~\ref{fig:snap1}, but for $\eta_{c}=0.01$. Only additional structures not found at $\eta_{c}=0.0034$ are shown.}
\label{fig:snap2}
\end{figure}

An overview of the different cluster structures found at the end of the simulation runs and in the experiment is presented in Figs.~\ref{fig:snap1} and~\ref{fig:snap2}. 
In particular, Fig.~\ref{fig:snap1} shows the simulation structures obtained at colloid packing fraction $\eta_{c}=0.0034$ for the SWY potential and $\sigma_{d}(0)=8 \, \sigma_{c}$.
We find clusters with sizes between $n_{c}=2$ (doublets not shown) and $n_{c}=9$, with $n_c$ the number of particles belonging to a cluster.  

In particular, we find that for  $n_c \leq 7$ the clusters have the same structures as Lennard-Jones  clusters~\cite{HOARE:1971p3888}. We find triplets for $n_c=3$ (Fig.~\ref{fig:snap1}a), tetrahedra for  $n_c=4$ (Fig.~\ref{fig:snap1}b), triangular dipyramids for  $n_c=5$ (Fig.~\ref{fig:snap1}c),  octahedrons for $n_c=6$ (Fig.~\ref{fig:snap1}d), and pentagonal dipyramids for  $n_c=7$ (Fig.~\ref{fig:snap1}e). For large number of particles we find the snub disphenoid for $n_c=8$ (Fig.~\ref{fig:snap1}f) and triaugmented triangular prism for $n_c=9$ (Fig.~\ref{fig:snap1}g, with two different orientations). 
In computer simulations  clusters with  particle numbers $n_c\geq10$ are obtained at higher packing fractions. The additional  structures obtained at colloid packing fraction $\eta_{c}=0.01$ are shown in Fig.~\ref{fig:snap2}. 
The square dipyramid is found for  $n_c=5$ (Fig.~\ref{fig:snap1}a)
Also, for the clusters with $n_c=10$ ( gyroelongated square dipyramid in Fig.~\ref{fig:snap2}b) and $n_c=12$ (icosahedron in Fig.~\ref{fig:snap2}d) we find good agreement with experiments. 
The cluster with $n_c=11$ (icosahedron minus one, Fig.~\ref{fig:snap2}c) was, on the other hand, not found in experiments. As this structure is identical to the icosahedron except for one missing particle, it can easily be  missed in the experimental FESEM micrographs. 

These structures are also in good agreement with those observed in previous experiments~\cite{Manoharan:2003p937,Wagner:2010p3885}. 
As noted by~\citet{Manoharan:2003p937}, clusters containing $n_c=8$ (snub disphenoid), $n_c=9$ (triaugmented triangular prism) and $n_c=10$ (gyroelongated square dipyramid) particles are members of a set of convex polyhedra~\cite{JOHNSON:1966p3889}. 

\begin{figure}
\includegraphics[width=4cm]{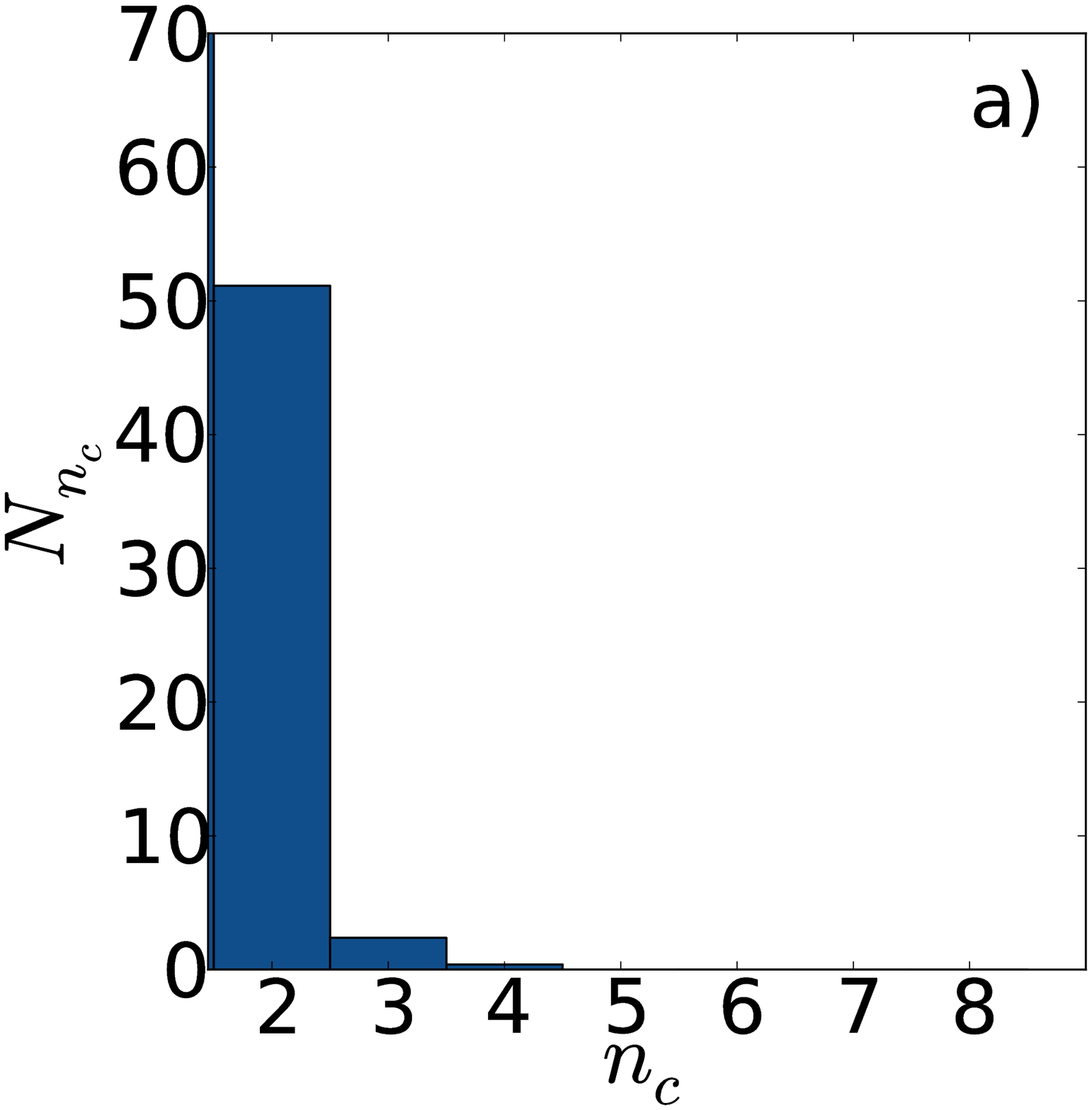}
\includegraphics[width=4cm]{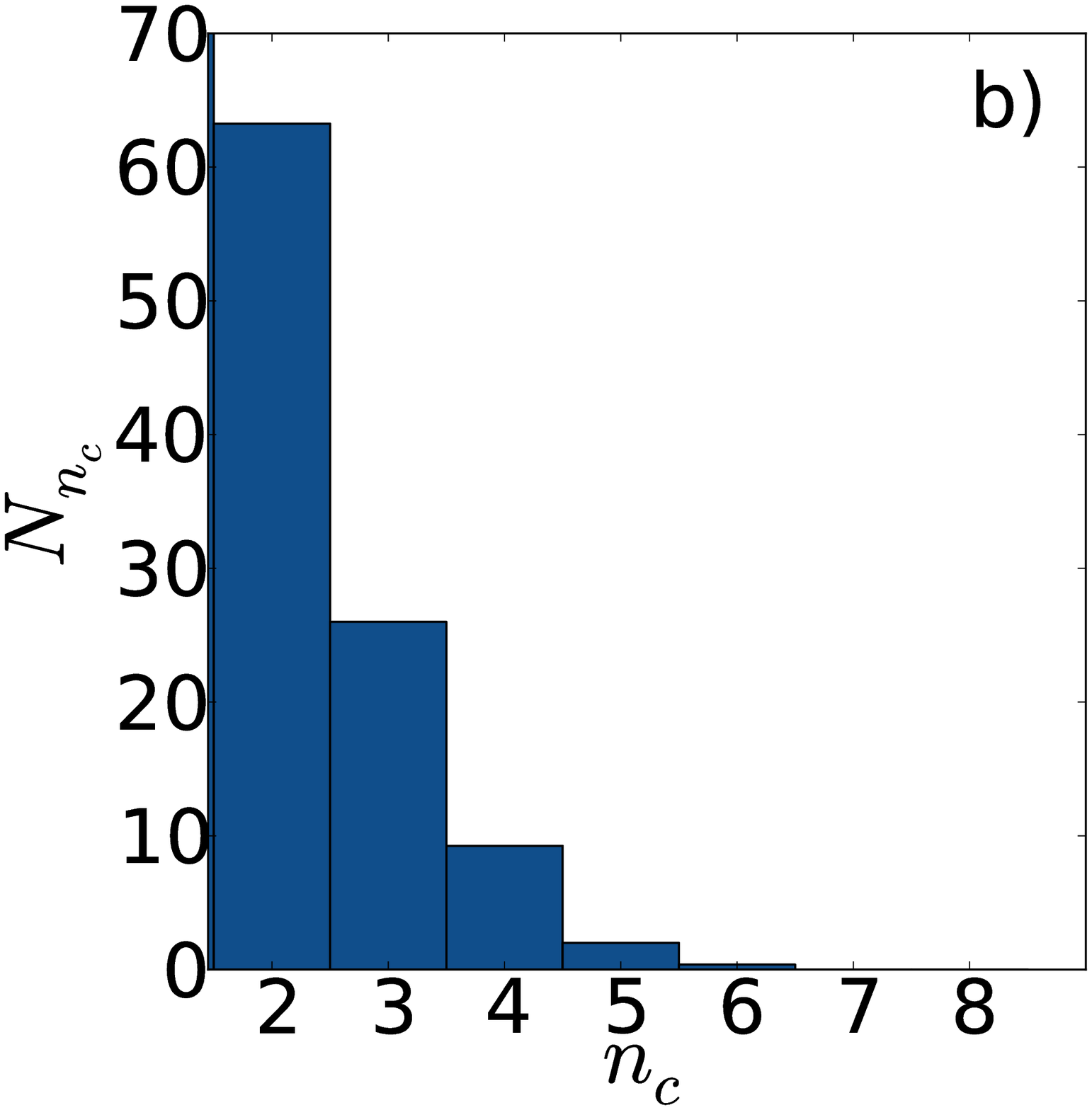}
\includegraphics[width=4cm]{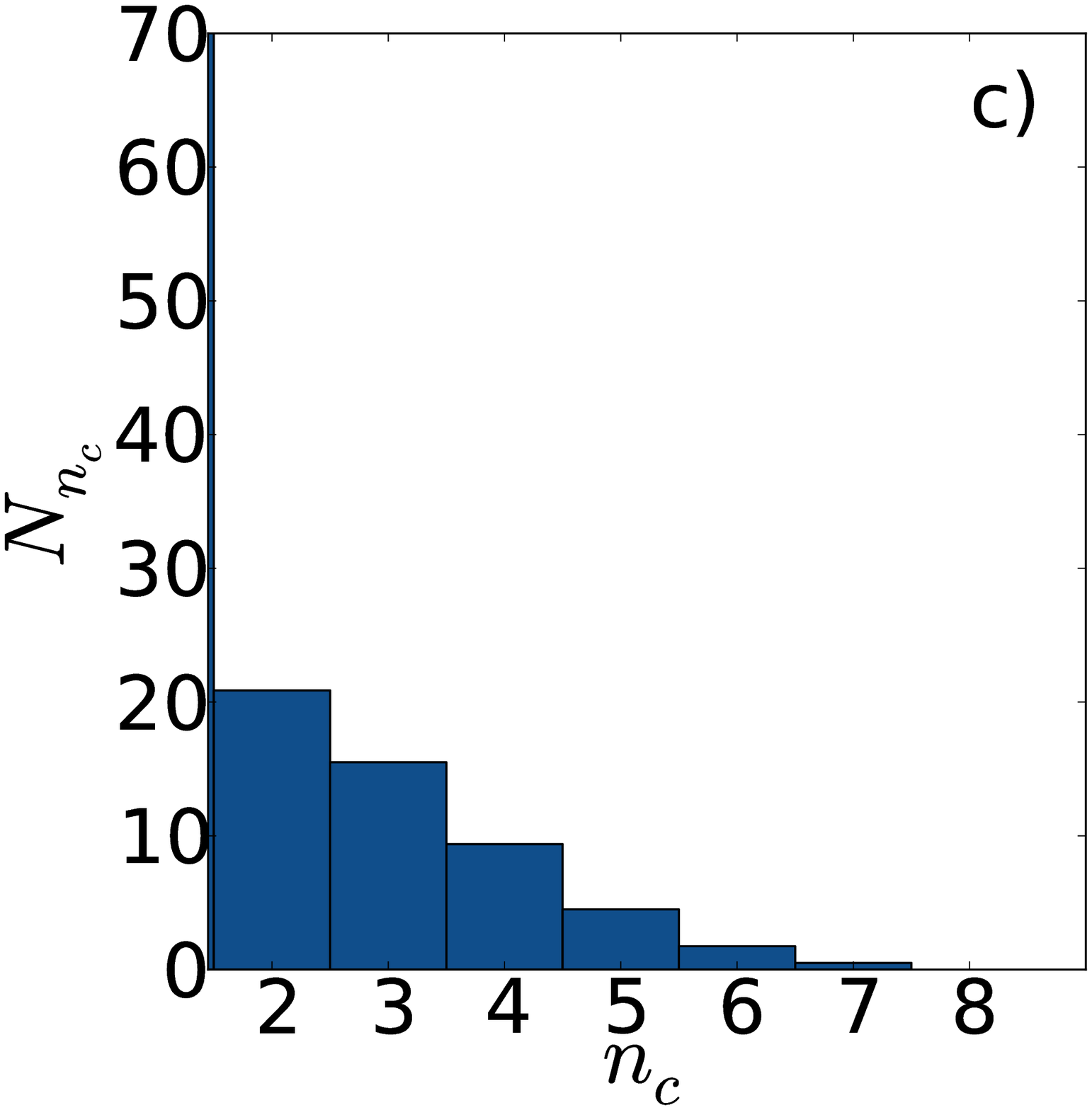}
\includegraphics[width=4cm]{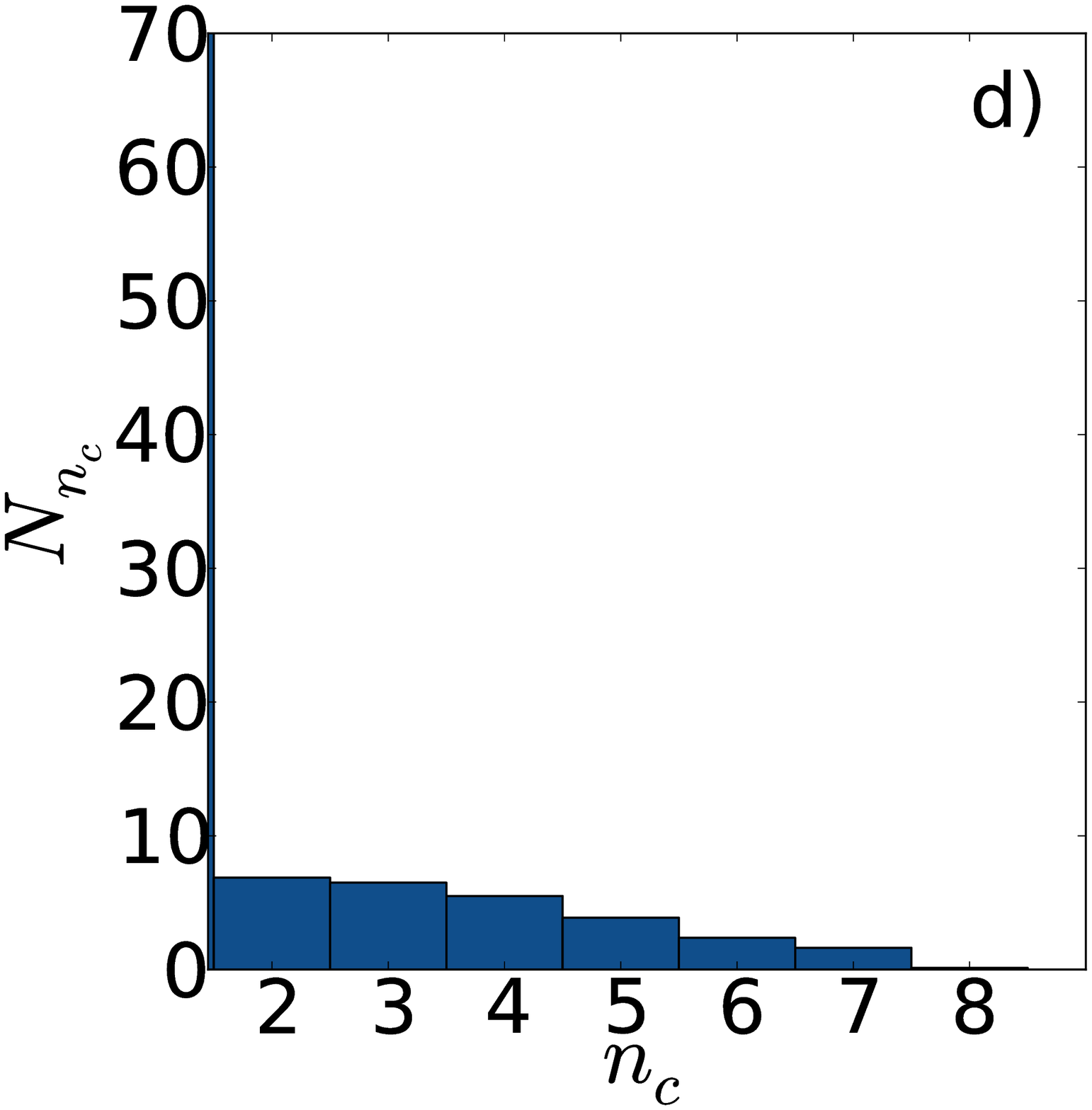}
\caption{Distribution of the number of clusters, $N_{n_{c}}$,  as a function of the number of particles in the cluster, $n_{c}$, found in computer simulations (SWY, $\eta_{c}=0.0034$). For clarity, the $n_{c}=1$ bar is omitted. Shown are result for different initial droplet sizes: a) $\sigma_{d}(0)/\sigma_{c}=2$. b) $\sigma_{d}(0)/\sigma_{c}=4$. c) $\sigma_{d}(0)/\sigma_{c}=6$. d) $\sigma_{d}(0)/\sigma_{c}=8$.}
\label{fig:hist}
\end{figure}

Larger cluster structures can be obtained not only by increasing the colloid packing fraction but also by increasing the starting droplet size, as demonstrated by the histograms of the number of clusters $N_{n_{c}}$, with $n_{c}$  the number of colloidal particles forming the cluster. Figure~\ref{fig:hist} shows the cluster distribution for varying starting droplet sizes $2< \sigma_{d}(0)/\sigma_{c}<8$. From Fig.~\ref{fig:hist}a) to  Fig.~\ref{fig:hist}d) the distribution becomes broader for larger droplet diameters $\sigma_{d}(0)$, while at the same time the yield  of smaller clusters decreases. 
The presence of a greater number of large clusters at larger droplet diameters can be explained by the larger surface available to capture colloidal particles in the initial stages of the simulation. Likewise, the probability to capture a small number of colloidal particles decreases with increasing droplet surface, which  leads to a decrease in the yield of small clusters. 
%The degree of polydispersity of the emulsions could
%  influence the experimental size distribution, but we do not expect the general trend to be different to the one
%  shown in Fig.~\ref{fig:hist} for the simulations.

\begin{figure}
\includegraphics[width=8cm]{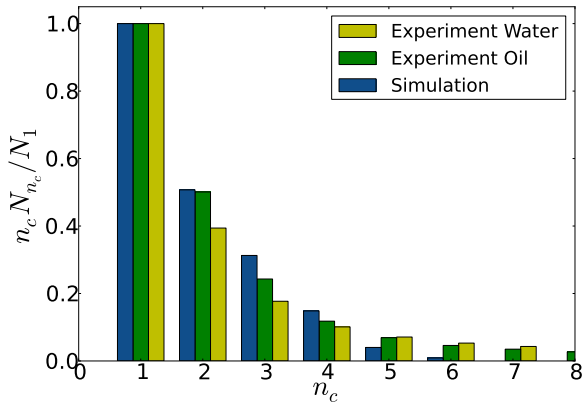}
\caption{Comparison between the number of particles $n_{c} N_{n_{c}}/N_{1}$ belonging to a cluster of size $n_{c}$ found in simulations and the weight fraction of particles as a function of $n_{c}$ measured in experiments (see Figures 4 and 6 of Ref.~\cite{Wagner:2010p3885}). The simulation results are for starting droplet size $\sigma_{d}/\sigma_{c}=4$, and $\eta_{c}=0.0034$. In the experiments the amount of building blocks added via the water or oil phase was 108 mg. For both simulation and experiments the percentage of particles per oil was 3.4\%.}
\label{fig:comp}
\end{figure}

The comparison with experiments can be made more quantitative by comparing the experimentally~\cite{Wagner:2010p3885} measured weight fraction of the clusters with the total number of particles, $n_{c} N_{n_{c}}$, belonging to a cluster with $n_{c}$ colloids. In order to compare the two quantities we normalize the experimental results with the weight fraction of single particles and the simulation results by the number of single particles $N_{1}$.
In the first experiments by Wagner et al, the particles were
  dispersed in the oil phase. In subsequent work \citet{Wagner:2010p3885}, experimentally compared this situation with that of adding particles via the water phase. They found that
  the same cluster structures result and that the cluster size
  distributions are also similar. In the present paper, for
  consistency, in Fig. 8 we compare the simulation results to
  experiments where the particles were added via the water phase. We
  keep, as an additional data set, the size distributions that were
  obtained by adding particles via the oil phase.
Figure~\ref{fig:comp} shows the two experimental results together with the simulation results  for the SWY potential, $\sigma_{d}(0)/\sigma_{c}=4$, and $\eta_{c}=0.0034$ corresponding to a percentage of particles per oil of 3.4\%.  
%We note how~\citet{Zerrouki:2006p4273} at  5\% particles per volume of oil, found a fraction with respect to singlets of 0.75, 0.7 and 0.47 for doublets, triplets and quadruplets, respectively.
%These are higher yields than what we found in our simulations and experiments. 
As the degree of polydispersity of our emulsions is low~\cite{Wagner:2010p3885}, we do not expect it to have a significant effect on the experimental size distribution, and the comparison to the (monodispersed) simulation results is viable. \citet{Zerrouki:2006p4273} ], at 5\% in weight of silica microparticles per volume of oil, found a fraction with respect to singlets of 0.75, 0.7 and 0.47 for doublets, triplets and quadruplets, respectively. Although these are higher yields than what we found in our simulations and experiments, their experimental parameters differ significantly from ours, so that no conclusion about the relative performance of both methods can be drawn.
%%%%%

In simulations, we find that clusters with the same number of constituent particles can still have a variety of different structures (isomers). 
Instead of distinguishing between all possible isomers we classify clusters based on their number of bonds. The bond-number  $n_{b}$ is defined as the total number of bonds in a cluster and, although unable to distinguish between all possible isomers,  gives an indication of the compactness of the cluster; for a given value of $n_{c}$ a smaller number of bonds indicates a more open structure as compared to a cluster with more bonds. 

\begin{figure}
\includegraphics[width=8cm]{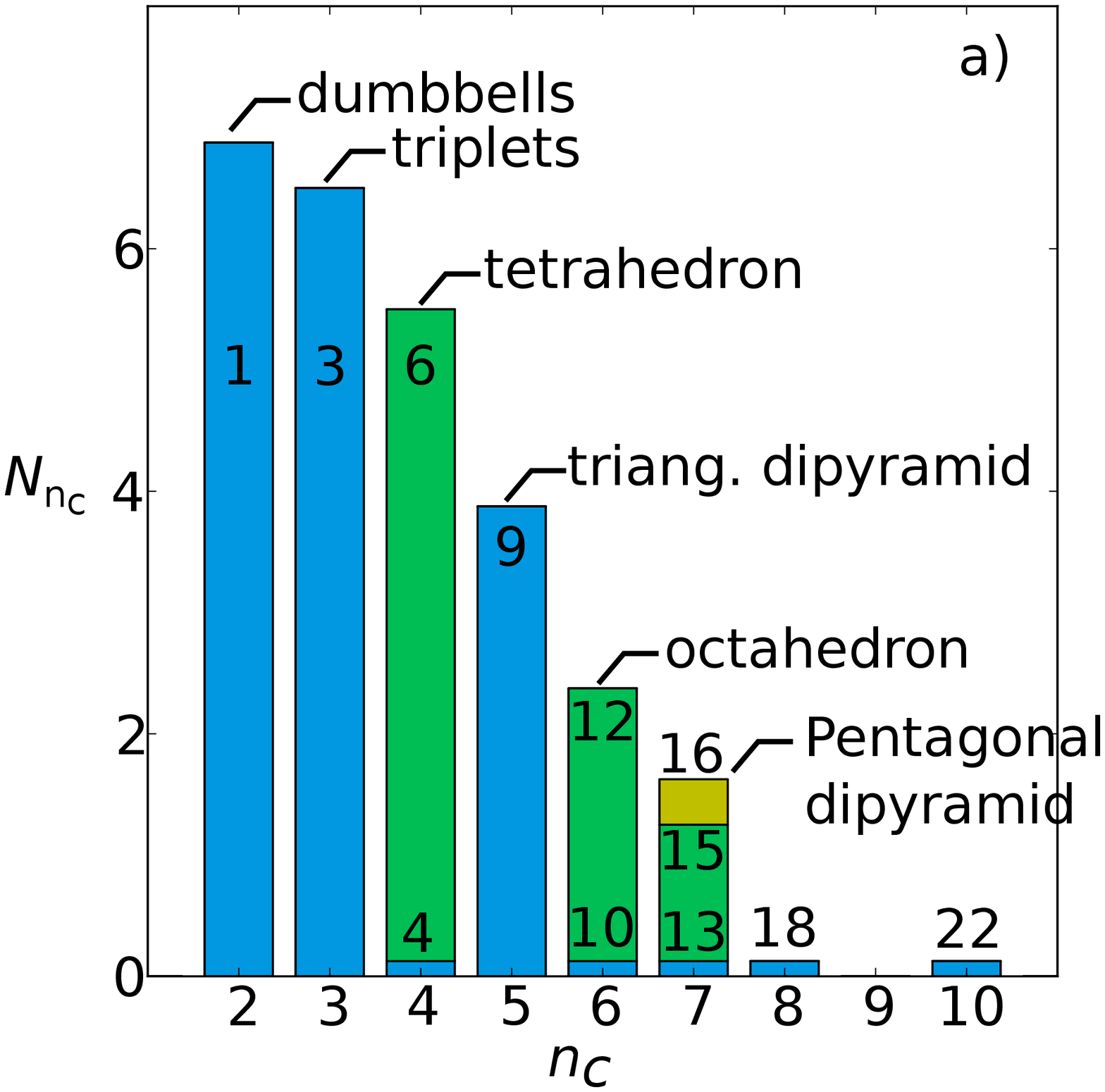}
\includegraphics[width=8cm]{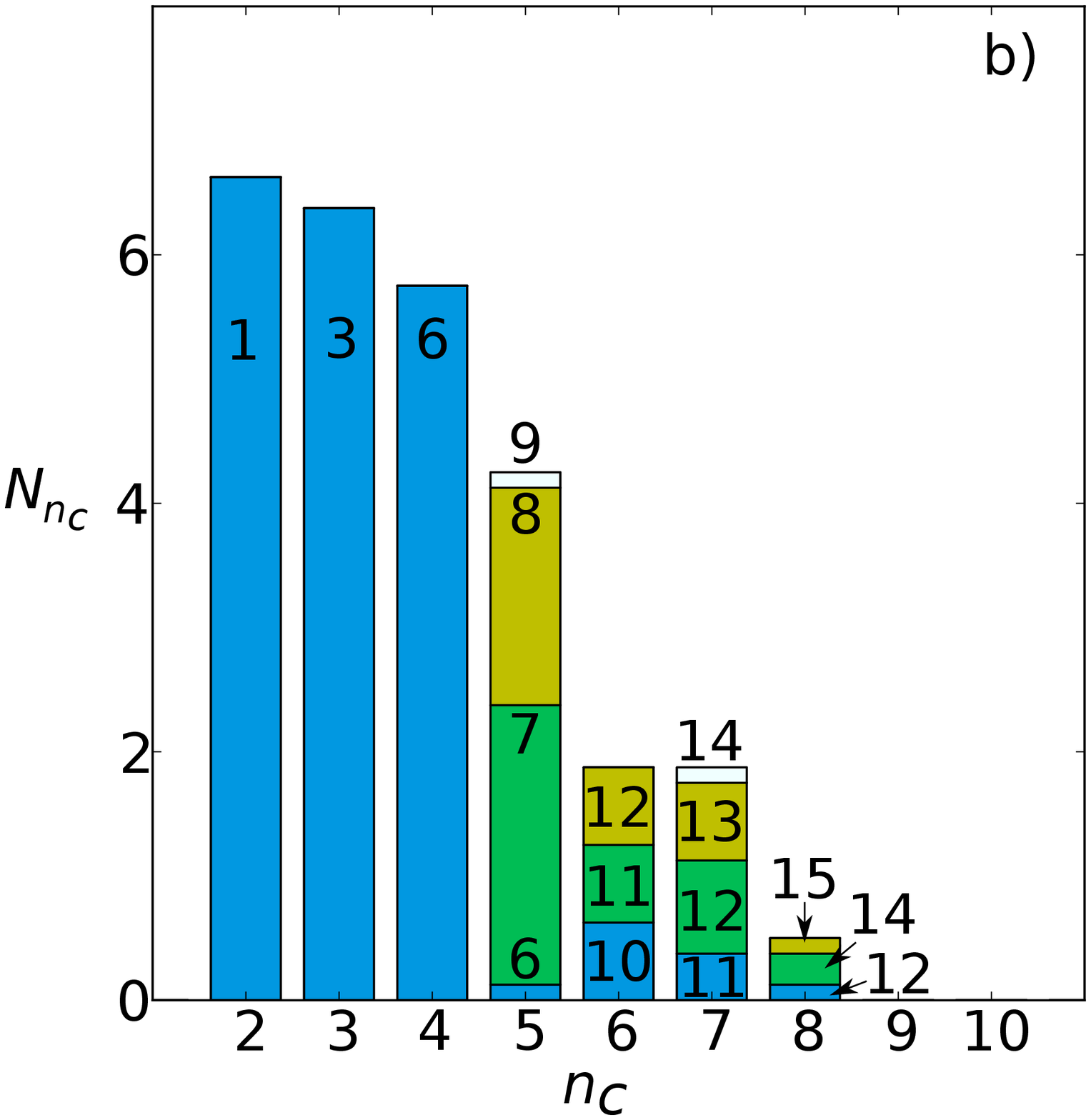}
\caption{Number of cluster $N$ with $n_{c}$ colloidal particles ($\sigma_{d}(0)=8 \, \sigma_{c}$ and $\eta_{c}=0.0034$). The total height of the columns indicates the number of cluster $N_{n_{c}}$ with $n_{c}$ particles. 
Each bar is divided in differently colored regions with a relative size proportional to the number of clusters with $n_{b}$ bonds. The numerical label indicate the bond-number for the region. For clarity, the $n_{c}=1$ bar is omitted.  a) Results for the SWY potential. Shown are also the names of the most relevant structures. b) Results for the AOY potential.}
\label{fig:links}
\end{figure}

Fig.~\ref{fig:links}a shows a stacked histogram of the number of clusters with a specific bond-number (SWY potential,  $\sigma_{d}(0)=8 \sigma_{c}$, $\eta_{c}=0.0034$).
The total height of the columns indicates the number of cluster $N_{n_{c}}$ with $n_{c}$ particles. 
Each bar is divided in differently colored regions with a relative size proportional to the number of clusters with $n_{b}$ bonds. Each region is labeled with the actual bond-number. 
For $n_{c}=2, 3$, only one type of cluster is found with $n_{b}=1, 3$, respectively. Clearly these bond-numbers correspond to doublets and triplets, respectively.
For $n_{c}=4$, two different structures are found, a small fraction of clusters with an open structure with only four bonds, and a structure with six bonds, corresponding to the tetrahedron shown in Fig.~\ref{fig:snap1}b.
For increasing number of constituent particles the number of isomers increases.
Fig.~\ref{fig:links}b) shows the stacked histogram of the number of clusters with a specific bond-number for the AOY potential  ($\sigma_{d}(0)=8 \sigma_{c}$, $\eta_{c}=0.0034$).
Strikingly, the AOY potential produces a larger number of different isomers than the SWY potential. 
In particular,  the AOY potential has isomers with smaller bond-numbers compared to the SWY potential. We interpret the more open structures that we find for the AOY potential as a direct result of its steep attraction and the resulting slow equilibration of the cluster geometry.

\subsection{Hierarchical assembly: Superclusters}
\label{s:super}
In order to investigate the possibility of hierarchical assembly of colloidal particles using the droplet-evaporation technique we carried out computer simulations of a mixtures of tetrahedral clusters and emulsion droplets. 
Hence, we prepared an initial configuration of the simulation consisting of clusters with tetrahedral symmetry (Fig.~\ref{fig:snap1}b). No single particles or other than tetrahedral cluster structures were present.

The tetrahedral clusters are thermal in the sense that the particles forming the clusters are kept together solely by the short-ranged attraction and can in principle dissolve, i.e. bonds can break on a long time scale by thermal activation. In experiments, bond breaking is even more unlikely because clusters are held together by van der Waals interactions, which are much stronger than the attraction used in our model. Nevertheless, structures that require bond-breaking in order to form are easily recognized. 

\begin{figure*}
\includegraphics[width=16cm]{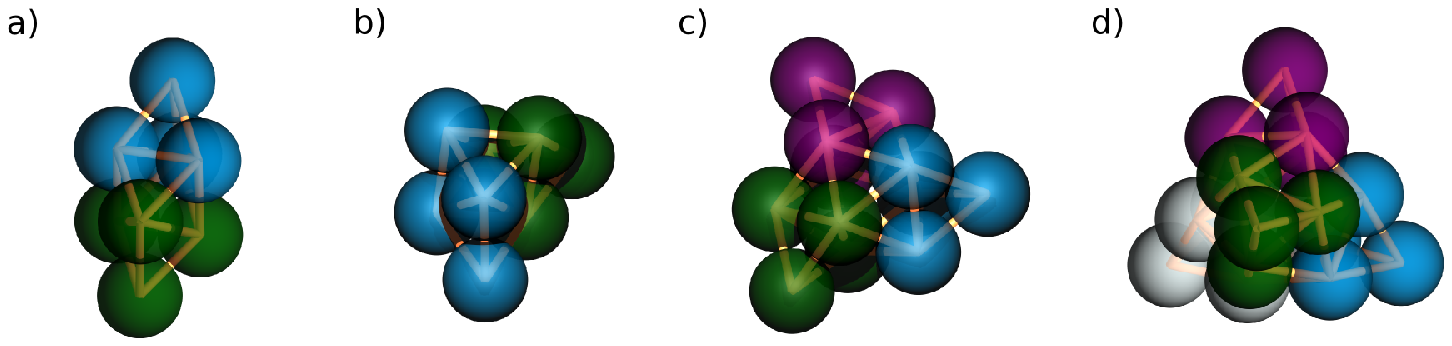}
\caption{Supercluster structures (SWY potential, $\sigma_{d}(0)=9 \sigma_{c}$, $\eta_{c}=0.1$). Particles with the same color belong to the same initial tetrahedral building block. a) Octahedral dipyramid, b) truncated hexagonal layers, c) Supercluster composed of three tetrahedra, d) Supertetrahedron. }
\label{fig:snap3}
\end{figure*}

Figure~\ref{fig:snap3} shows the structures obtained in the simulations (SWY potential, $\sigma_{d}=9 \sigma_{c}$, $\eta_{c}=0.1$).
In particular, Fig.~\ref{fig:snap3}a-b) show two structures consisting of two tetrahedral building blocks. The octahedral dipyramid (Fig.~\ref{fig:snap3}a) is formed by two tetrahedra rotated by $30^{o}$ against each other and with two faces touching. Fig.~\ref{fig:snap3}b) shows two truncated hexagonal layers. This cluster  formation is possible only because one of the initial tetrahedral clusters has dissolved. This structure is therefore not accessible experimentally when non-thermal clusters are used.  
Fig.~\ref{fig:snap3}c) shows superclusters of three tetrahedra, while Fig.~\ref{fig:snap3}d) shows a  supertetrahedron, i.e. a cluster formed by four tetrahedra arranged  a tetrahedral geometry.

\section{Summary and Conclusions}
\label{s:conc}
We investigated cluster formation via emulsion droplet evaporation with computer simulations and experiments.
We used Metropolis Monte Carlo simulations to model the process of cluster formation in a binary mixture of colloidal particles and emulsion droplets. The colloidal particles interact via both a short-ranged attraction and a long-ranged repulsion,  while the second component that represents the emulsion droplets interacts only with an attractive well with the colloids. This potential well has a minimum at the droplet surface in order to induce the Pickering effect. 
The droplet-droplet interaction is a hard-core interaction with a hard-sphere diameter chosen so that the droplets cannot merge.  The droplets shrink at a fixed rate, in order to model experimental conditions of droplet evaporation. 

We also performed experiments on polystyrene spheres 154 $nm$ in diameter in a toluene-water emulsion. 
The emulsion was vitrified and analyzed with cryo-FESEM, before the evaporation process. The micrographs indicated a random distribution of the positions of the particles that are trapped at the droplet surface. These results can be surprising because charged particles act as electric dipoles when trapped onto droplets due to the effect that the part of their surface exposed to the nonpolar solvent (toluene) cannot sustain its charge. The resulting long-ranged dipolar repulsions  may result in regular spatially separated arrangements of the particles at the interface~\cite{nikolaides2002,horozov2003}. For micron-sized particles this is supported by optical micrographs that indicate polyhedral arrangements when a small number of particles are bound onto an emulsion droplet~\cite{manoharan2006}. A different scenario can be expected for our submicron-sized particles because their higher diffusivity can interfere with repulsive interactions, and suppress regular orientation. Interestingly, an earlier study of droplets stabilized by a large number of submicron-sized colloids could even demonstrate that the particles are not necessarily kept separated from each other and can form close-packed islands or even a monolayer~\cite{binks2002}.

Computer simulation snapshots and radial distribution functions were used to analyze the dynamics of cluster formation in computer simulation. In agreement with experimental results we find that in our model the particles can freely diffuse on the surface  of the droplet before the evaporating droplets force particle agglomeration into clusters.
The degree of ordering of the particles on the droplet surface depends on the range of the repulsive interaction, (Debye screening length)  that in our model is of the order of two particle diameters. Choosing a longer range for the repulsion could lead to ordered distribution of particles on the surface of the droplets. Furthermore the strength of dipolar interactions, neglected in our present model, could be relevant.

 %%%%%

After the complete evaporation of the droplets we find stable clusters that range from sphere doublets  to complex polyhedra. The structures and size distributions found in simulations matched those found in experiments. 
Histograms show that larger clusters can be obtained by increasing the initial size of the droplets or by increasing the density of colloidal particles at the expense of a smaller yield for smaller clusters in accordance with the results of~\citet{Wagner:2010p3885}.

The bond number was used to distinguish different structures with the same number of constituent particles. 
We found that although the AOY potential gives the same clusters and size distributions as the SWY potential, the AOY interaction results in a larger number of possible structures than the SWY interaction.
In particular, the AOY potential gives structures with a smaller number of bonds, i.e. with more open structure. This is intuitively reasonable, since the steep attractive part of the AOY potential results in a difficult equilibration of the geometric structure of the clusters.  
Although these potentials do not model quantitatively our experimental system, this study can give an indication of what type of interactions one should use in order to change the cluster morphology to more open (softer) clusters.

Our simple model reproduces the experimental results accurately despite a lack of realistic energy or time scales. It is therefore sensible to assume that the model captures the essential physics of the assembly process and that more complex assembly processes can be studied with a certain confidence. Hence, the model can be useful  to guide experimental work. As an example we applied the theoretical model to a fluid mixture of tetrahedra clusters and droplets. The computer simulations show that the assembly process via emulsion droplet evaporation can lead to stable superclusters with two, three and four tetrahedral building blocks. These novel structures are not found in the assembly of single nanoparticles and could represent a step in the direction of novel and complex mesoscale materials. 

We neglected in our theoretical model the formation of dipole moments for particles trapped at the droplet surface. Including dipolar interactions and increasing the range of the repulsion both constitute interesting steps beyond the current work. Furthermore, using our present model for studying gelation, as was recently reported in colloidal dispersions with a small immiscible liquid~\cite{Koos:2011p4276}, could be interesting.

\acknowledgments We thank  Douglas J. Ashton and Helmut R. Brand for useful discussions. We thank Beate F\"orster and Martina Heider, Bayreuth Institute of Macromolecular Research, for taking cryo-FESEM images. I.S. acknowledges the Elite Network of Bavaria for support in the frame of the Elite Study Program "Macromolecular Science". We acknowledge the DFG for financial support via SFB840/A3.

%\bibliography{clusterbib}
%\bibliographystyle{aipnum4-1}

\end{document}